%% file: main.tex
\documentclass[twocolumn]{aastex631}

\usepackage{graphicx}	
\usepackage{amsmath}	

\newcommand{\bp}{MWBP}
\newcommand{\mws}{MWS}
\newcommand{\Teff}{T$_{\rm eff}$}

\begin{document}

\title[DESI-MWS Backup Program]{The Backup Program of the Dark Energy Spectroscopic Instrument's Milky Way Survey}

\correspondingauthor{Arjun Dey}
\email{arjun.dey@noirlab.edu}




\input{author_list}



\begin{abstract}

The Milky Way Backup Program (MWBP), a survey currently underway with the Dark Energy Spectroscopic Instrument (DESI) on the Nicholas U. Mayall 4-m Telescope, works at the margins of the DESI Main surveys to obtain spectra of millions of additional stars from the {\it Gaia} catalog. 
Efficiently utilizing twilight times ($<18^\circ$) and poor weather conditions, 
the \bp\  extends the range of stellar sources studied to both brighter magnitudes and lower Galactic latitude and declination than the stars studied in DESI's Main Milky Way Survey. 
While the \bp\ prioritizes candidate giant stars selected from the {\it Gaia} catalog (using color and parallax criteria), it also includes an unbiased sample of bright stars (i.e., $11.2 \lesssim G < 16$~mag) as well as fainter sources (to $G \lesssim 19$~mag). As of March 1, 2025, the survey had
obtained spectra of $\sim 7$ million stars, 
approximately $1.2$ million of which are included in the DESI Data Release 1. 
The full survey, when completed, will cover an area of more than 21,000~deg$^2$ and include approximately 10 million {\it Gaia} sources, roughly 
equal to the number of stellar spectra obtained through the DESI Main Survey, while only utilizing $<9\%$ of all DESI observing time. 
\end{abstract}

\keywords{Sky Surveys --- Redshift Surveys --- Stellar kinematics --- Giant Stars --- Milky Way Galaxy}

\newcommand\jn[1]{\textcolor{purple}{#1}}

\section{Introduction}

The Dark Energy Spectroscopic Instrument \citep[DESI;][]{DESI_Instrument_2022,DESIExp_Instrument_2016} on the Nicholas U. Mayall 4m Telescope of the Kitt Peak National Observatory is undertaking a 5-year mission with the primary goal of measuring the cosmological parameters that define the accelerating expansion history of the universe \cite[e.g.,][]{DESI_BAO_III_2024,DESI_BAO_IV_2025,DESI_Cosmology_VI_2024,DESI_Cosmology_VII_2024,DESI_DR2_Cosmology.2025}. The major portion of the cosmological survey is carried out during dark time, when the contribution of moonlight to the sky background is minimal and 
redshifts of faint, distant galaxies can be measured with high precision \citep{Schlafly2023}. 
During bright time, the project targets
a nearby bright galaxy sample \citep{DESI_BGS_2023} and, in addition, 
measures the radial velocities of 
millions of 
stars in the Milky Way (MW). While 
this latter bright-time survey (the Main Milky Way Survey, hereinafter \mws) is primarily aimed at measuring the kinematics of the stellar halo of the MW, it will also yield stellar parameters (\Teff, $\log{g}$, [Fe/H], [$\alpha/$Fe], and some elemental abundances) for several million stars \citep{Cooper2023}.

These dark-time and bright-time surveys 
(which we refer to here as the DESI Main Survey) occupy all the ``good" observing conditions, when the sky is mostly clear and the sky brightness permits these primary survey observations to be carried out efficiently. However, for a significant fraction of time, the conditions are too poor to carry out the main program, but the dome can remain open and the telescope and instrument are usable for observations.  
These conditions can occur during morning or evening twilights
(i.e., typically when the sun is between $\sim12^\circ$ and $18^\circ$ below the horizon); 
at times when there is no threat of rain, but the sky is very cloudy or changeable; or when the combination of moonlight and thin clouds makes the sky brightness too high to carry out the Main Survey. But even these worst of times can result in scientifically productive data. 
During these periods, we carry out the ``Milky Way Survey Backup Program" (hereinafter \bp). 
%
The \bp\ works in the interstices of the DESI Main survey: the DESI observing team tracks the weather conditions and determines when to switch to the \bp\ and when the Main Survey observations can recommence.  
%

In making efficient use of poor weather conditions, the \bp\ 
complements the 
\mws\
and significantly increases the total number of stellar targets observed.
Firstly, the \bp\ covers a larger footprint than the 
\mws, extending to lower Galactic latitude. Secondly, the MWBP extends to 
brighter stellar targets. Thirdly, the MWBP sparsely samples fainter stars, down to {\it Gaia} magnitudes of $G\sim 19$ mag. Because the observations are taken under poor conditions, the spectra tend to be of lower quality than the MWS observations of similar magnitude stars. Nevertheless, spectra from the DESI \bp\ constitute $\approx22\%$ of the stellar spectra in the DESI Data Release 1 \citep{DESI_DR1_2025}, and up to a third of the stars in the future second data release (see \S~\ref{sec:DataAvailability} for further details). 

\begin{figure}[ht]
\centering
\includegraphics[width=0.45\textwidth]
{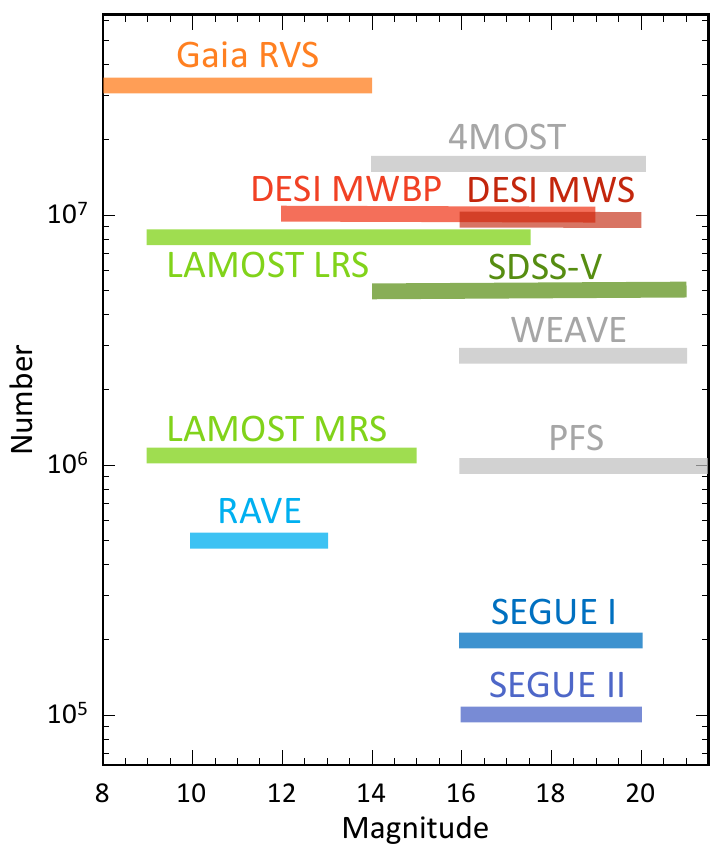}
\caption{Comparison of the magnitude range and size of recent, ongoing (colored bars), and future (gray bars) stellar spectroscopic surveys with those of the \bp, which is among the largest stellar spectroscopic surveys currently underway or planned. The \bp\ complements the \mws\ in extending to brighter stars. Both surveys measure spectra for stars fainter than the {\it Gaia} RVS limit and are each likely to include $\sim 10$~million stellar spectra when complete.  }
\label{stellar_survey_comparison}
\end{figure}

As of March 1, 2025, roughly 8.7\% of the available observing time has been used for the backup program, 
producing approximately 7 million stellar spectra.
We expect that the \bp\ will eventually deliver $\sim$ 10 million stellar spectra, a data set comparable in size to the MWS and among the largest stellar spectroscopic data sets to date. 
Figure~\ref{stellar_survey_comparison} compares the anticipated size and magnitude range of the \bp\ to those of other recent (and future) spectroscopic surveys of $> 10^5$ stars carried out with spectral resolution comparable to or greater than that of DESI. The parameters for the other surveys are taken from Table 1 of \citet[]{Cooper2023} with two exceptions and several additions. Firstly, the number of stars in LAMOST LRS is updated with values from its DR10 v2.0 (released 29 September 2024; \url{https://www.lamost.org/lmusers/}), which includes derived stellar parameters for 7.4 million AFGK stars 
and 0.9 million M-type giants, dwarfs, and subdwarfs. In addition, the magnitude range for RAVE is converted to {\it Gaia} G. Secondly, rough parameters are shown for SDSS-V \citep{Almeida23} and the future surveys 4MOST \citep{Chiappini19}\footnote{See also \url{www.4most.eu/cms/science/gasconsurv}.}, WEAVE \citep{Jin24}, and PFS \citep{Takada14}. The relevant magnitude range is in {\it Gaia} G for SDSS-V, 4MOST, and WEAVE, and V-band for PSF.

In this paper, we describe the \bp\
in detail to help potential users understand the properties of these data and the differences from the 
\mws. We describe the \bp\ Survey target selection (\S\ref{sec:targetselection}), the tiling strategy (\S\ref{sec:tilingstrategy}), the observational strategy (\S\ref{sec:obsstrategy}), and the data quality (\S\ref{sec:dataquality}), illustrating 
the latter with some examples. We summarize the current state of the survey in section \S\ref{sec:summary}. 

\section{Target sample definition}
\label{sec:targetselection}

\begin{figure}[t]
\centering
\includegraphics[width=\linewidth]{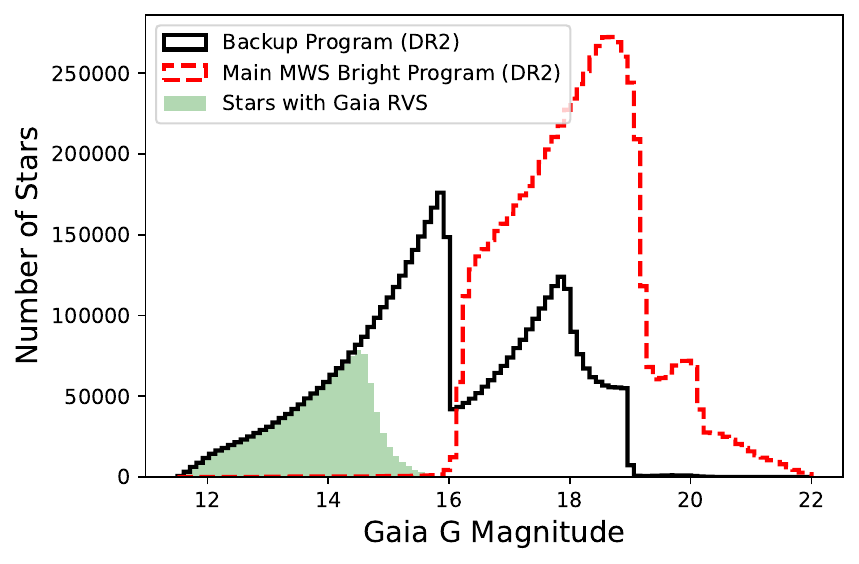}
\caption{The {\it Gaia} $G$-band magnitude distributions of the spectroscopically observed targets in the \bp\  (black histogram) and the  \mws\ (red dashed histogram). The subset of stars with radial velocity measurements from {\it Gaia} RVS are shown by the shaded green histogram. The DESI histograms are constructed from the targets observed during (approximately) the first three years of survey operations and which will be included in the second DESI Data Release (DR2). \label{fig:Gaia_mag_distribution}}
\end{figure}

The targeting pipeline for all DESI targets is described in detail by \citet{Myers2023}.
The primary targets for the \bp\ are bright stars selected from the Gaia DR2 catalog \citep{Gaia2016,Gaia2018b} across a larger footprint than the MWS \citep[see \S~4.1.4 of][for details]{Myers2023}. The 
\bp\ sample is divided into five mutually exclusive classes of candidates (see Table~\ref{tab:targetselection} for details): 
bright stars with $G<16$~mag ({\tt BACKUP\_BRIGHT}); 
candidate halo giants with $16 < G \lesssim 19$, divided into high and low priority targets based on parallax error ({\tt BACKUP\_GIANT} and {\tt BACKUP\_GIANT\_LOP});
and two filler samples of fainter backup stars ({\tt BACKUP\_FAINT} which contains stars $16\le G < 18$, and {\tt BACKUP\_VERY\_FAINT} which extends this by an additional magnitude $18\le G < 19$). 
Targets for all classes were selected to be north of declination $\delta=-30^\circ$ to be accessible from the 
Mayall Telescope, with {\it Gaia} $G$, $BP$, and $RP$ all fainter than $\approx$11.2~mag to avoid the possibility of saturating the DESI spectrum. The saturation limit as a function of $BP-RP$ color was determined from the DESI observations obtained during the Survey Validation phase \citep{Schlafly2023,DESI_EDR_2024}. 
The magnitude distribution of the \bp\ stars is complementary to that of the \mws, as shown in Figure~\ref{fig:Gaia_mag_distribution}. 

The $G<16$~mag {\tt BACKUP\_BRIGHT} candidate sample contains about 46~million stars with excellent astrometric data from the {\it Gaia} satellite. Among targets with $G\ge 16$~mag, priority is given to candidate halo giant stars, selected based on their {\it Gaia} parallax $\pi_{Gaia}$ and parallax uncertainty $\sigma_\pi$. 
The {\tt BACKUP\_GIANT} sample varies from about 80 stars per square degree at the North Galactic Pole (NGP) to about 220 stars per square degree at $\vert b \vert = 30^\circ$. 
The \bp\ subsamples the lower Galactic latitude {\tt BACKUP\_GIANT} sources at the rate of 
Min($10^{(2-\gamma)},1$), where:
\begin{eqnarray}
\gamma = {\rm log_{10}}~\rho_0\  {\rm if}\ {\rm log_{10}}\rho_0 < 0 \\
\gamma = 1.1{\rm log_{10}}\rho_0\  {\rm if}\ {\rm log_{10}}\rho_0 > 0
\end{eqnarray}
and where 
\begin{equation}
\begin{aligned}
    {\rm log_{10}}\rho_0 &= 6.9529434 + 0.18786695\ {\rm cos}(l) - 2.6621607  \\
    & \times\ {\rm log}_{10} \left[ \vert b\vert - 3.05095354\ {\rm cos}(l) + 8.10060927\right]
\end{aligned}
\end{equation}
is an approximation to the variation of stellar density with Galactic longitude ($l$) and latitude ($b$) ($\rho_0$ in units of stars per square degree).
Sources excluded by this subsampling are added to the {\tt BACKUP\_GIANT\_LOP} candidate class. Finally, the {\tt BACKUP\_FAINT} and {\tt BACKUP\_VERY\_FAINT} target samples contain a large set of fainter stars in a region 
that
excludes the Galactic plane according to the prescription described in Table~\ref{tab:targetselection}.

The final \bp\ target samples consist of 
45.6M, 1.85M, 26.1M, 25.0M, and 33.4M
stars in the {\tt BACKUP\_BRIGHT, BACKUP\_GIANT, BACKUP\_GIANT\_LOP, BACKUP\_FAINT and BACKUP\_VERY\_FAINT} target classes respectively. These target classes and the overall selection for the \bp\ are  summarized in Table~\ref{tab:targetselection}. These five 
candidate classes
are targeted 
in the following prioritized order: 
    {\tt BACKUP\_GIANT} $>$ {\tt BACKUP\_BRIGHT} $>$ {\tt BACKUP\_GIANT\_LOP} $>$ {\tt BACKUP\_FAINT} $>$ {\tt BACKUP\_VERY\_FAINT}.


\begin{table*}
\centering
\caption{Selection criteria for the five target classes of the \bp.}
\begin{tabular}{|l|l|}
\hline
\textbf{Selection} & \textbf{Criteria} \\ \hline
\multicolumn{2}{|l|}{\textbf{Cuts shared by all of the BACKUP classes: }} \\
\ \ \ Sky coverage: & $\delta$ $\geq -30^\circ$  \\ 
\ \ \ Finite color: & ${(BP-RP) \ne NaN}$ \\ 
\ \ \ Bright limit to avoid saturation: &  ${G \ge 11.2}$~mag\\
    & $BP \ge 11.2$~mag AND $RP \ge 11.2$~mag  \\ 
\hline
\multicolumn{2}{|l|}{\textbf{Primary sample (BACKUP\_BRIGHT):}} \\ 
\ \ \ Faint limit  & $G < 16.0$~mag \\ 
\ \ \ Bright limit & $G \ge 11.2 + 0.6(BP-RP)$ \\ 
\ \ \ Galactic latitude limit & $|b| > 7$ \\ 
\hline
\multicolumn{2}{|l|}{\textbf{High priority halo giant sample (BACKUP\_GIANT):}} \\ 
\ \ \ Faint limit & $G < {\rm min}(17.5 + 0.6(BP-RP), 19)$ \\
\ \ \ Bright limit & $G \ge 16.0$~mag \\ 
\ \ \ Parallax & $\pi_{Gaia} < 0.1~{\rm mas} + 2\sigma(\pi_{Gaia})$ \\ 
\ \ \ Galactic latitude limit& $|b| > 7$ \\ \hline
\multicolumn{2}{|l|}{\textbf{Low priority halo giant sample (BACKUP\_GIANT\_LOP):}} \\
\ \ \ Faint limit  & $G < {\rm min}(17.5 + 0.6(BP-RP), 19)$ \\
\ \ \ Bright limit & ${G \geq 16.0}$~mag \\
\ \ \ Parallax & ${\rm \pi_{Gaia} < 0.1 mas + 3\sigma(\pi_{Gaia})}$ \\ 
\ \ \ Not in BACKUP\_GIANT & \\ 
\ \ \ Galactic latitude limit & $|b| > 7$ \\ \hline
\multicolumn{2}{|l|}{\textbf{Filler sample (BACKUP\_FAINT)}:} \\
\ \ \ Faint limit  & $G < 18.0$~mag \\ 
\ \ \ Bright limit & $G \ge 16.0$~mag \\
\ \ \ Not in BACKUP\_GIANT & \\
\ \ \ Not in BACKUP\_GIANT\_LOP& \\ 
\ \ \ Exclude Galactic plane & (($b>-0.139l + 25^\circ$ AND $l\le180^\circ$)\\
   & OR ($b>0.139l - 25^\circ$ AND $l>180^\circ$) \\
   & OR ($b<-0.139l + 25^\circ$ AND $l>180^\circ$) \\
   & OR ($b<0.139l - 25^\circ$ AND $l\le 180^\circ$)) \\ \hline
\multicolumn{2}{|l|}{\textbf{Filler sample (BACKUP\_VERY\_FAINT)}:} \\
\ \ \ Faint limit  & $G < 19.0$~mag \\ 
\ \ \ Bright limit & $G \ge 18.0$~mag \\
\ \ \ Not in BACKUP\_GIANT & \\
\ \ \ Not in BACKUP\_GIANT\_LOP& \\ 
\ \ \ Exclude Galactic plane & (($b>-0.139l + 25^\circ$ AND $l\le180^\circ$)\\
   & OR ($b>0.139l - 25^\circ$ AND $l>180^\circ$) \\
   & OR ($b<-0.139l + 25^\circ$ AND $l>180^\circ$) \\
   & OR ($b<0.139l - 25^\circ$ AND $l\le 180^\circ$)) \\ \hline
\end{tabular}
\label{tab:targetselection}
\end{table*}

\section{Tiling Strategy}
\label{sec:tilingstrategy}
\begin{figure*}[ht]
\centering
\includegraphics[width=0.49\linewidth]{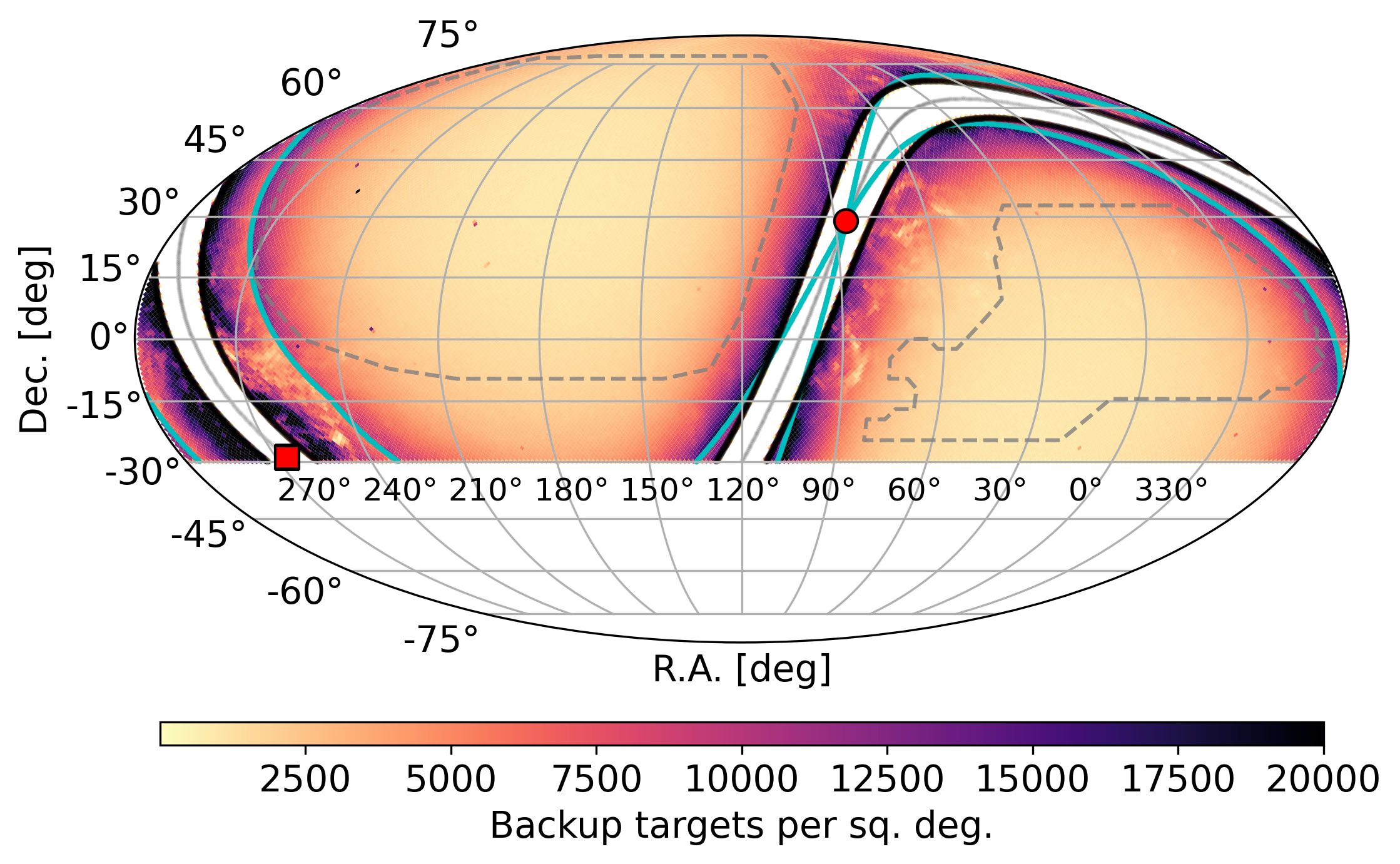}
\includegraphics[width=0.49\linewidth]{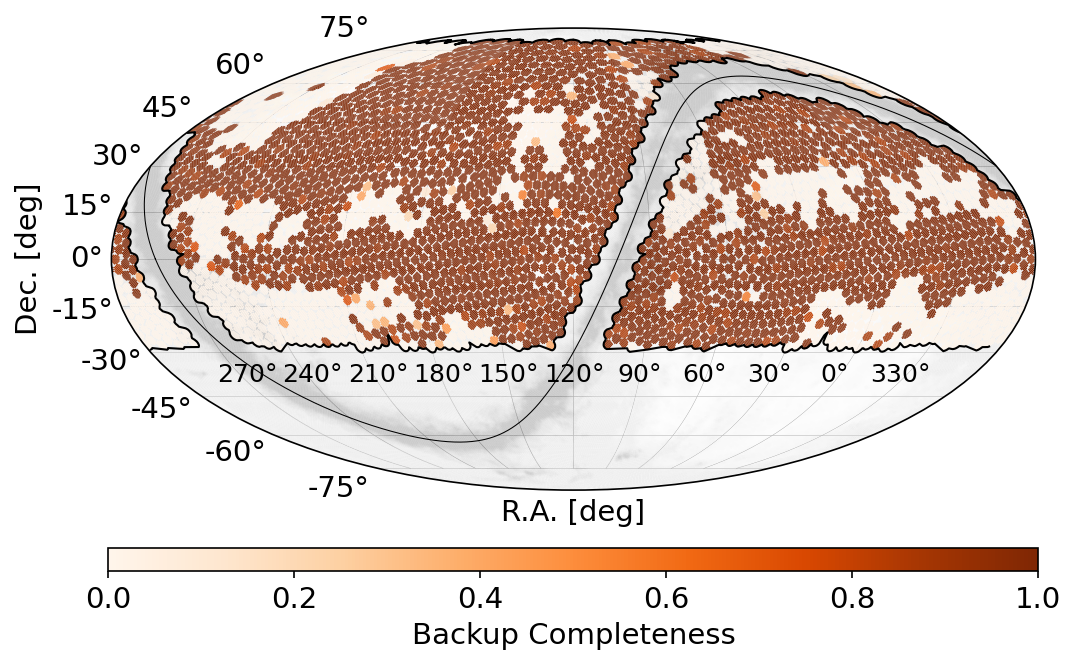}
\caption{Left: The footprint of the \bp. The cut at Galactic latitude $\vert b \vert\approx 7^\circ$ represents the boundary of the \bp\ footprint for the \texttt{BACKUP\_BRIGHT}, \texttt{BACKUP\_GIANT} and \texttt{BACKUP\_GIANT\_LOP} target classes, and the cyan solid lines show the variable $b$ limits of the \texttt{BACKUP\_FAINT} and \texttt{BACKUP\_VERY\_FAINT} selections (see Table~\ref{tab:targetselection}). The footprint of the \bp\ 
reaches to much lower Galactic latitude (roughly $\vert b \vert \gtrsim 7^\circ$) than the MWS, which is shown by the grey dashed lines. The sky density of all potential MWBP targets is shown by the color scale. The Galactic center and anti-center are marked by a red square and circle respectively.
Right: 
Areal completeness of the \bp\ as of 1 March 2025. The color bar shows the completeness as a fraction of the nominal exposure time of $t_{\rm eff} = 60s$; the red (and darker) dots 
represent tiles where the depth is at least 85\% of the nominal depth. 
Nearly 70\% of the tiles defined for the \bp\ have been observed to this depth. The jagged black contour represents the true boundary of the edge tiles, and therefore the expected coverage of the footprint once the \bp\ is complete.
\label{fig:backup_footprint}
}
\end{figure*}

DESI is capable of simultaneously obtaining spectra of targets distributed over a 3.2~deg field of view \citep{DESI_Corrector.Miller2024,DESI_Fiber_System.Poppett2024}.
DESI defines a ``tile'' as a single pointing of the instrument at a particular location on the sky, with a fixed set of targets. 
Details of the tiling strategy for the DESI main survey have been presented by \citet{Schlafly2023}. 
The footprint for the \bp\ is an expanded version of the DESI footprint, extending south to declination $\delta\ge-30^\circ$ and covering 
the range of Galactic latitudes described in Table~\ref{tab:targetselection} (see also Figure~\ref{fig:backup_footprint}). 
The \bp\ uses a tiling of 2657 tiles to cover this entire expanded footprint of more than 21,000 deg$^2$. This tiling is equivalent to the first layer (`Pass 0') of the DESI Main Survey, but extends over the larger footprint of the \bp. 
By design the \bp\ tiles do not overlap each other (see Figure~\ref{fig:backup_footprint}). The full tiling, when completed, will result in spectra of $\approx$10 million {\it Gaia} sources (compared with the $\approx$9 million stars observed by the end of the
\mws).

The fiber assignment for a given tile is done on the fly just prior to its observation \citep[see \S~5.6 of][]{Schlafly2023}. Briefly, for each tile, sources are selected from the \bp\ target catalogs and prioritized as described in the previous section. 
Since \bp\ targets range over a wider footprint than the DESI Main dark- and bright-time surveys, extending beyond the region covered by the Legacy Surveys imaging \citep[see][]{DESI_LS_Imaging}, standard stars for \bp\ tiles are selected from the {\it Gaia} catalog as described in \S~4.2 of \citet{Myers2023} (i.e., the {\tt GAIA\_STD\_BRIGHT, GAIA\_STD\_FAINT} target classes).  If there are any target-of-opportunity (TOO) triggers, these are also added during fiber assignment, typically with higher priority than the \bp\ targets. This process typically results in approximately 4000 \bp\ targets per tile, a number which varies with Galactic latitude, with the highest latitude tiles having as few as $\approx3000$ science targets per tile. Unlike tiles for the DESI Main bright survey (where targeting is prioritized for galaxy targets from the Bright Galaxy Survey), \bp\ tiles contain almost exclusively stellar targets. TOO triggers are rare, with the entire DESI DR1 (DR2) release only containing 6 (98) such targets. 

Although candidate targets are selected across the entire footprint using the criteria defined in Table~\ref{tab:targetselection}, the tile selection is further restricted to tiles with centers satisfying the Galactic latitude limit $\vert b\vert \gtrsim 7^\circ$ and the declination range $-28.5 < \delta < 80$. This results in targets in the top three priority categories (i.e., {\tt BACKUP\_GIANT}, {\tt BACKUP\_BRIGHT}, and {\tt BACKUP\_GIANT\_LOP}) being included over the full footprint down to $\vert b \vert \gtrsim 7^\circ$, but the two filler samples (i.e., {\tt BACKUP\_FAINT} and {\tt BACKUP\_VERY\_FAINT}) being further restricted by the longitude-dependent Galactic latitude cut described in Table~\ref{tab:targetselection}. 

\section{Observations \& Strategy}
\label{sec:obsstrategy}
 
The observational strategy for all parts of the DESI survey is described in detail in \cite{Schlafly2023}. Here, we briefly describe the parts relevant to the triggering of the \bp\ and the actual range of conditions that result from the strategy. 

\subsection{Observing Speed and Triggering Strategy}

During a night when the instrument is on sky and observing, the observing software continuously computes the ``survey speed'', a dimensionless quantity 
that
is related to the efficiency of observing a given tile under the current conditions. DESI defines the survey speed $S$ 
as 
the rate of accumulation (per elapsed second) of effective exposure time,
where the effective exposure time is 
the time required to reach the required signal-to-noise ratio (SNR) in ``nominal'' photometric, zenith-observing conditions of 1.1~arcsec seeing, $r$-band sky brightness of 21.07 AB mag per square arcsec, and no extinction by Galactic dust. $S$ can vary between zero (when the observation of a given tile is completely prevented due to clouds) to $\approx2.5$ under excellent conditions. 
The \bp\ is executed during twilight (typically between $\sim12^\circ$ and $18^\circ$ twilight, i.e., before it is dark enough for DESI Main Survey observations) 
and during times of poor weather (i.e., when the transparency, seeing, sky brightness, or some combination of these conditions are poor). The \bp\ observations are triggered by software, with only occasional overrides by the Lead Observing Scientist (the individual at the telescope who is responsible for the observations on any given night) when $S\le 0.08$. 

\begin{figure}[ht]
    \centering
    \includegraphics[width=\linewidth]{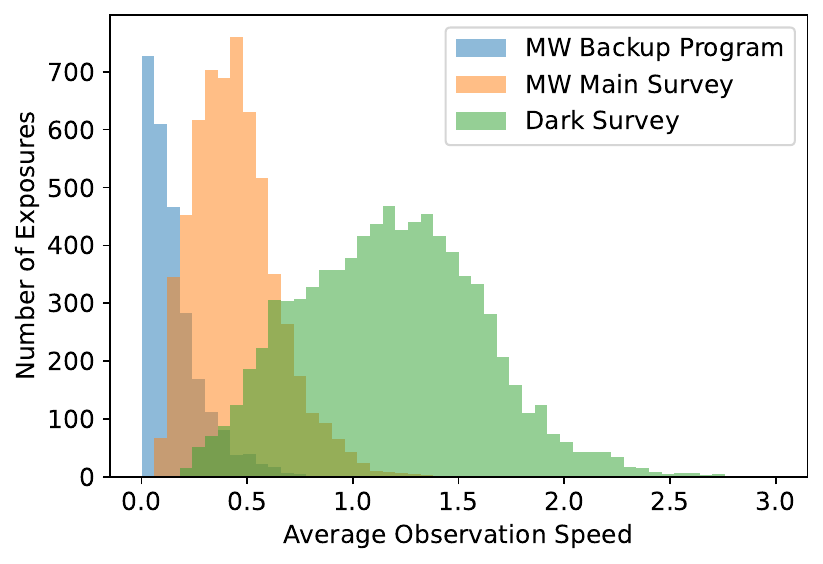}
    \caption{Distribution of average observing speed (see text) for the \bp\ (Blue histogram), \mws\ (orange histogram) and the DESI Main dark-time cosmology survey (green histogram) exposures. }
    \label{fig:speeddistribution}
\end{figure}

While $S$ is not recorded for each spectrum, the average speed during the observation of a given source can be reconstructed from the DESI catalog data as follows: 
\begin{equation}
    \bar{S} = t_{\rm eff}\times f_{EBV} \times X^{1.75}/t_{exp}
\end{equation}
where, $t_{\rm eff}$ is the effective exposure time computed by the DESI Exposure Time Calculator during the observing, defined as the time required to reach the required signal-to-noise ratio in ``nominal'' photometric, zenith-observing conditions of 1.1~arcsec seeing, $r$-band sky brightness of 21.07 AB mag per square arcsec, and no extinction by Galactic dust; $f_{EBV}=10^{1.732E(B-V)}$ is the scaling factor applied for the Galactic extinction at the location of the tile; $X$ is the airmass at the time of observation, and $t_{exp}$ is the actual elapsed exposure time \citep[see][for details]{Schlafly2023}.\footnote{For all tiles observed after October 22, 2024, the term $f_{EBV}$ was dropped from the observing speed estimate used at the telescope when exposing on \bp\ tiles. This will slightly decrease the resulting $t_{\rm eff}$ for observations after this date, especially in regions of high Galactic extinction, but does not affect any observations included in the DESI DR1 and DR2 releases.}
Figure~\ref{fig:speeddistribution} shows the resulting distribution of the average observing speed for the \bp\ and a comparison to the observing speed distributions for the DESI Main bright-time Milky Way Survey and the DESI Main dark-time cosmology survey \citep[see also Figure 6 of][]{Schlafly2023}. 

\begin{figure}[ht]
\centering
\includegraphics[width=\linewidth]{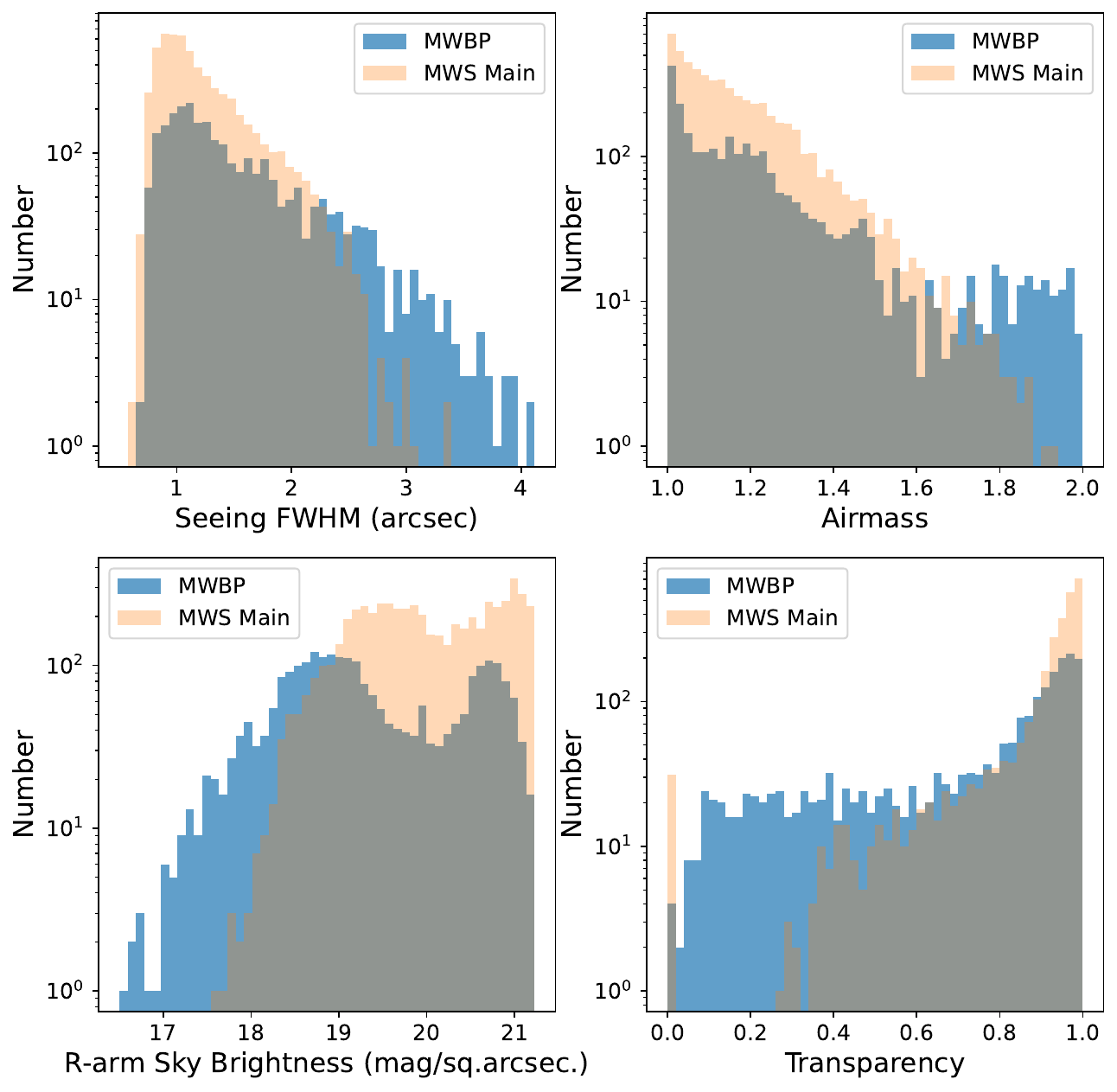}
\caption{Distribution of sky conditions for the exposures taken for the \bp\
(blue) and the 
\mws\
(orange). The four panels (clockwise from top left) show the seeing full-width at half maximum, airmass, transparency and $r$-band sky brightness.
}
\label{observingconditions}
\end{figure}

Since low observing speeds can result from a wide range of conditions, the \bp\ observations can be triggered for a variety of reasons. 
Figure~\ref{observingconditions} shows the distribution of the observing conditions during which \bp\ observations have been made. Although the \bp\ observations span a wide range, most \bp\ observations are triggered in conditions of either poor seeing (FWHM$\gtrsim2$~arcsec), poor transparency ($\lesssim0.3$), bright sky ($\gtrsim$19~mag/sq.arcsec.), or high airmass ($\gtrsim1.5$), or a combination of these. 


\subsection{Resulting Exposure Times}

Tiles for the \bp\ are observed with an 
effective exposure time goal of $t_{\rm eff}=60$~sec.
Since the conditions can be quite poor (and perhaps rapidly varying) for the \bp, the range in effective time for any given exposure can vary significantly. Some exposures only result in advancing this time by 1 sec (!) of effective time, whereas a few exposures may actually result in deeper observations than needed. Typically, many exposures are needed to achieve the goal time.

\begin{figure*}[t]
\centering
\includegraphics[width=\linewidth]{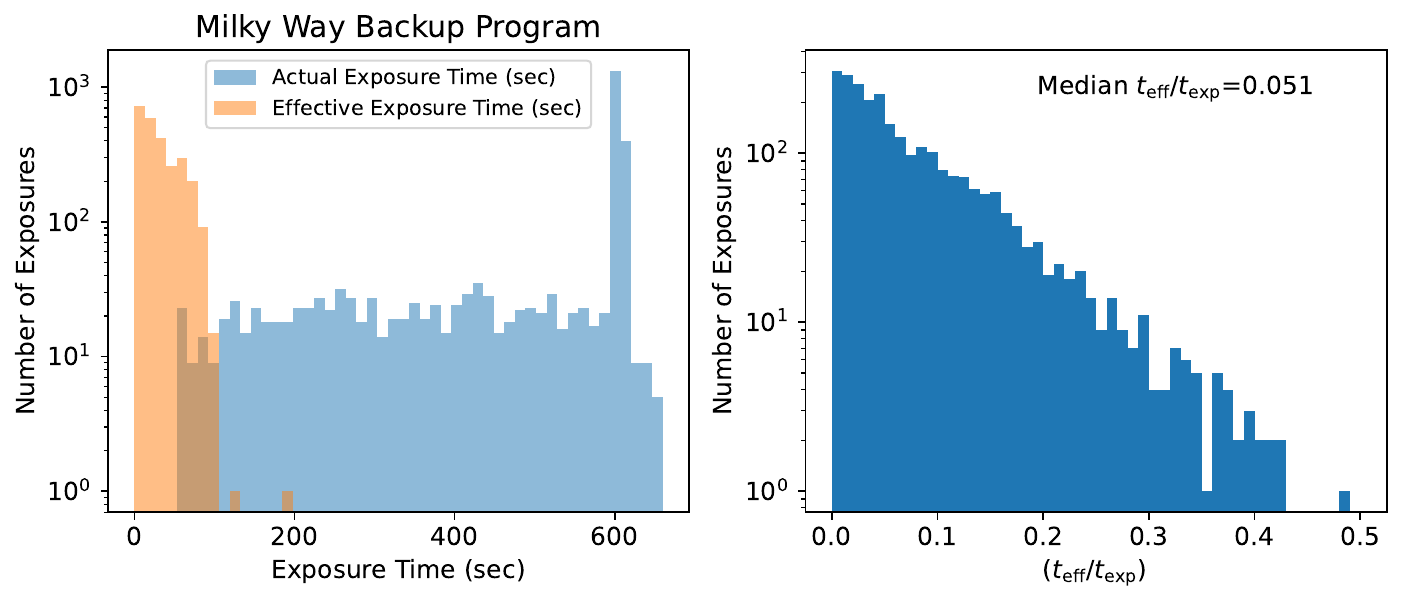}
\caption{Left: Distribution of the actual on-sky exposure time (blue) and the effective exposure time (orange) for the 
\bp\ exposures.
Right: The distribution of the ratio of the effective exposure time ($t_{\rm eff}$) to the actual elapsed time ($t_{\rm exp}$) for each observation.}
\label{fig:exptimes}
\end{figure*}

The resulting distribution of exposure times is shown in Figure~\ref{fig:exptimes}. Compared to the goal of 
$t_{\rm eff}=60$~sec, the bulk of the exposures have $t_{\rm eff}\ll 60$~sec, with a small number of exposures being over-exposed (due to the conditions changing rapidly during the exposure). However, 
actual exposure times per observation of more than $\sim 10$ minutes
are typically needed 
to achieve this effective exposure time distribution, with the median value of the ratio of $t_{\rm eff}/t_{exp} \approx 0.051$. 

\begin{figure}[ht]
\centering
\includegraphics[width=\linewidth]{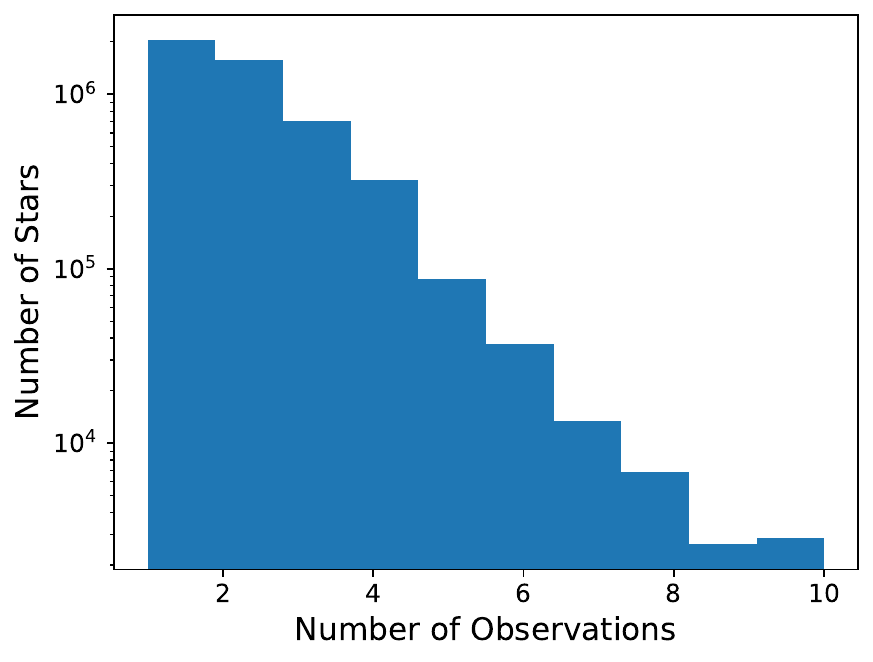}
\caption{Distribution of the number of observations per target. Nearly 60\% of the \bp\ stars have more than 1 observation.}
\label{fig:repeats}
\end{figure}

The strategy results in many repeat observations for individual sources until at least 85\% of this exposure time ($\approx 51$~sec) is achieved. The distribution of the number of observations per tile achieved in the \bp\ is shown in Figure~\ref{fig:repeats}.

\section{Data from the \bp}
\label{sec:dataquality}

\subsection{Current status}

The \bp\ observations obtained thus far 
represent only about 8.7\% of all of the observing time used by DESI. As of February 12, 2025, DESI has observed nearly 1821 ($\approx 68\%$) of the planned 2657 \bp\ tiles to at least 85\% of their nominal depth (i.e., $t_{\rm eff}\gtrsim 51$~sec). 

\subsection{Data Availability}
\label{sec:DataAvailability}

\begin{figure*}[ht]
    \centering
    \includegraphics[width=0.49\linewidth]{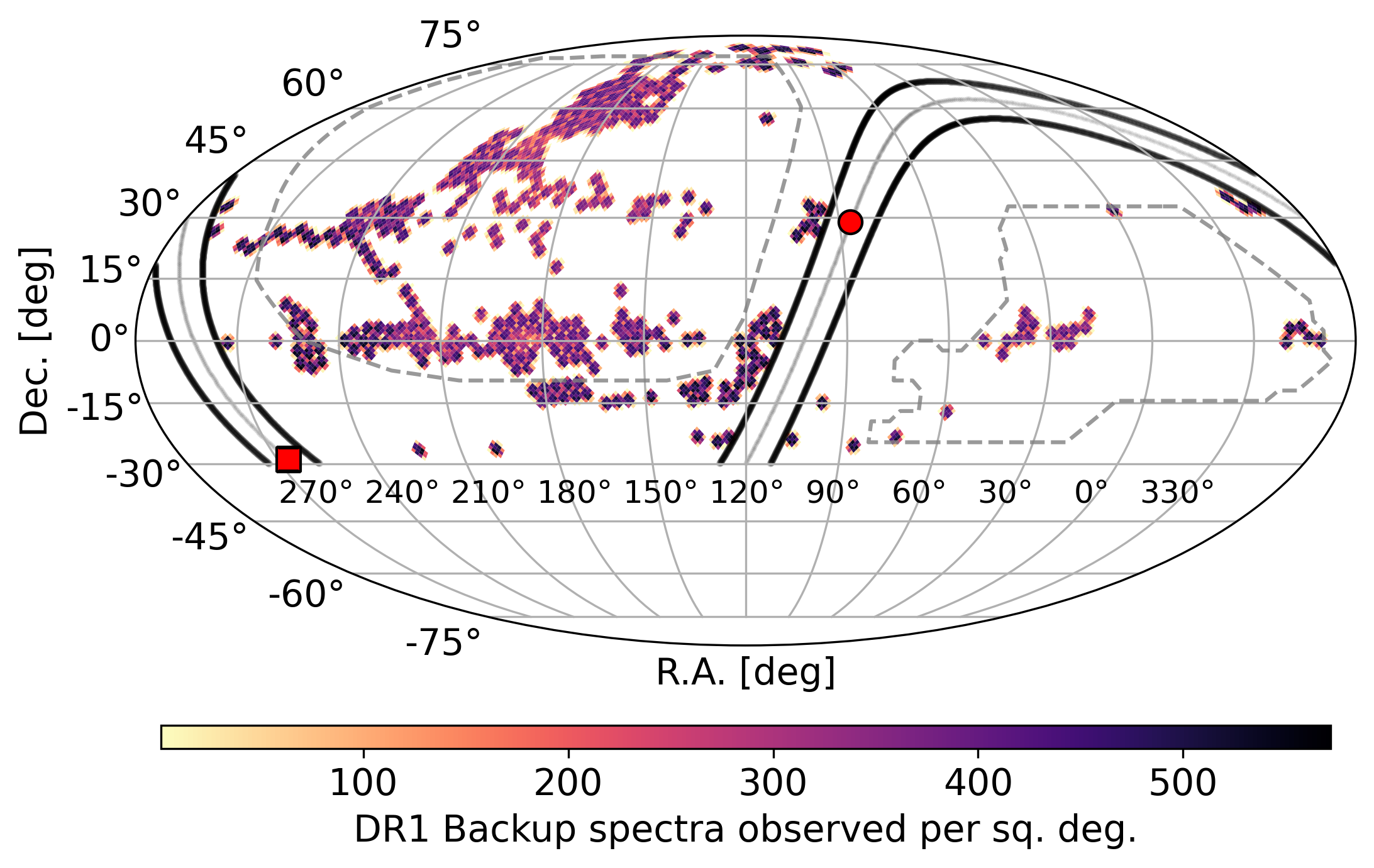}
    \includegraphics[width=0.49\linewidth]{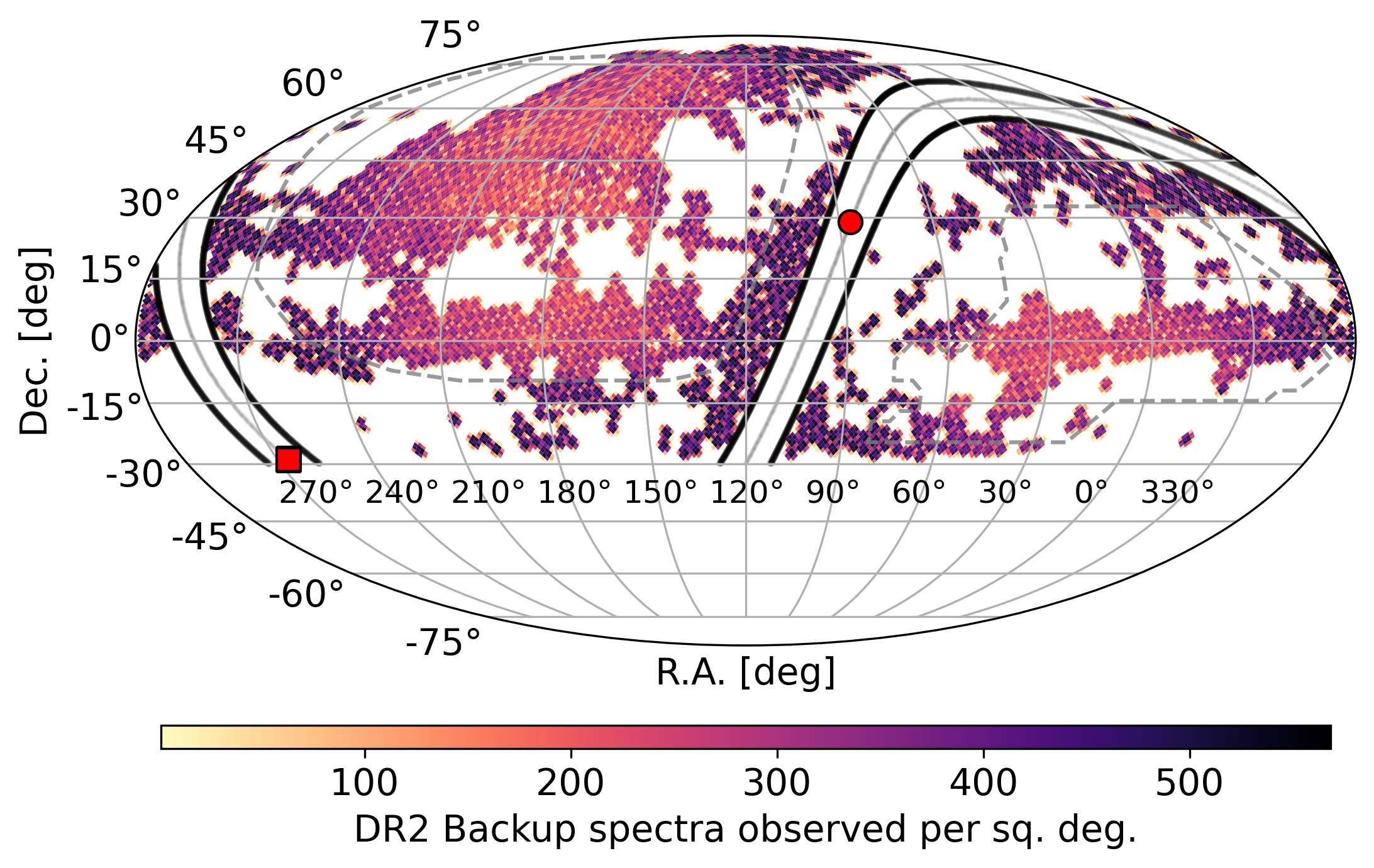}
    \caption{The sky density and distribution of \bp\ spectra in the DESI DR1 dataset (left) and the future DR2 dataset (right). The grey dashed line shows the slightly smaller footprint covered by \mws. The Galactic center and anticenter are marked by a red square and circle respectively and the solid black lines represent the \bp\ Galactic latitude limit of $\vert b \vert = 7^\circ$.}
    \label{fig:backup_dr1_dr2}
\end{figure*}

The DESI project schedules periodic public releases of the spectroscopic data and associated value added catalogs. The data from the \bp\ have been included as part of the DESI Early Data Release (EDR) and Data Release 1 (DR1). 
EDR contains 148k \bp\ spectra 
taken during DESI Survey Validation phase and the One Percent Survey \citep{DESI_EDR_2024}. DR1 includes 1.2M \bp\ spectra obtained during the first thirteen months of DESI operation \citep{DESI_DR1_2025}. 
Data Release 2 (DR2), which will encompass the first 3 years of DESI survey spectroscopy and will contain 4.9M spectra of \bp\ targets, is scheduled for release by early 2027. The DR1 release represents spectra collected from 327 unique tiles (i.e., $\approx12\%$ of the total footprint, or $\approx$3000~deg$^2$); the DR2 release will increase this to 1310 tiles (i.e., $\approx 49\%$ of the footprint). We refer the reader to \cite{DESI_EDR_2024,DESI_DR1_2025} for the details of the specific data releases. The sky coverage of the \bp\ in the DR1 and DR2 releases is shown in Figure~\ref{fig:backup_dr1_dr2}. The number of spectra observed in each Target Class within each DESI Data Release is shown in Table~\ref{tab:target_class}. 

\begin{deluxetable*}{lrrrrr}
    \tablecaption{Target class distribution across different data releases. \\\label{tab:target_class}}
    \tablehead{
        \colhead{Target Class} & \colhead{All Targets\textsuperscript{a}} & \colhead{Tile Targets\textsuperscript{b}} & \colhead{DR1} & \colhead{DR2} & \colhead{Total\textsuperscript{c}}
    }
    \startdata
        BACKUP\_BRIGHT & 45,677,684 & 22,472,468 & 636,850 & 2,832,416 & 6,260,000 \\
        BACKUP\_GIANT & 1,853,482 & 1,382,372 & 117,882 & 464,857 & 1,000,000 \\
        BACKUP\_GIANT\_LOP & 26,127,761 & 17,245,747 & 148,239 & 573,094 & 1,220,000 \\
        BACKUP\_FAINT & 24,958,299 & 18,406,150 & 196,848 & 664,914 & 1,210,000 \\
        BACKUP\_VERY\_FAINT & 33,392,580 & 24,001,223 & 116,580 & 369,782 & 640,000 \\
        \hline
        Total \bp\ Targets & 132,009,808 & 83,507,960 & 1,216,399 & 4,905,063 & 10,330,000\\
        \hline
        \mws\ & & 30,470,033 & & & 9,890,000 \\
    \enddata
    \tablecomments{\textsuperscript{a}All targets at declination $> -30$ degrees. \textsuperscript{b}Targets covered by the \bp\ tiles. \textsuperscript{c} Approximate number of stars anticipated at the completion of the survey, based on mock fiber assignment for the full \bp\ tiling.}
\end{deluxetable*}


 The \bp\ targets can be identified within any of these data releases by selecting sources according to  the tags
{\tt SURVEY="main"} and {\tt PROGRAM="backup"}. 
For comparison, the \mws\ targets can be found selecting {\tt SURVEY="main"} and {\tt PROGRAM="bright"}. Targets for each of the five target class described in Table~\ref{tab:targetselection} can be identified using the targeting bit values [58, 59, 60, 61, 62] recorded in the {\tt MWS\_TARGET} tag for the [{\tt BACKUP\_GIANT\_LOP,
BACKUP\_GIANT,  BACKUP\_BRIGHT,  BACKUP\_FAINT,  BACKUP\_VERY\_FAINT}] target classes, respectively \citep[see also Appendix B4 of][]{DESI_DR1_2025}. The DESI \bp\ spectra constitute $\approx 22\%$ (33\%) of the 
\mws\ sources in the DESI DR1 (DR2) release.

In what follows, we provide a broad description of the quality of the \bp\ data, separating out, where useful, the spectra released in March 2024 as part of DR1, and the spectra that will be included in the DR2 release. The pipeline processing associated with DR1 is described in full detail by \citet{DESI_DR1_2025}; the processing associated with DR2 will be described in a future paper. For additional details regarding the spectra of stars in DR1, please see the detailed description of the DESI Stellar Catalog (released as the DR1 MWS Value-Added Catalog) in \cite{Koposov2025}\footnote{The DESI DR1 Stellar Catalogue is available at \url{https://data.desi.lbl.gov/doc/releases/dr1/vac/mws/}.}.

\subsection{Calibrations and Reductions}

The data are reduced using the standard DESI spectroscopic pipeline described in \citet{DESI_SpecPipeline_2023}. Given the wide range of (typically poor) observing conditions during which \bp\ spectra are obtained, individual exposures can vary significantly in their SNR. When the median SNR$<2$ for the standard stars in a given exposure, the pipeline does not attempt to derive a full reconstruction of the throughput per wavelength element from the standard stars in the observation, but instead reverts to using a predefined calibration vector normalized to the broad-band flux of the stars. While this mode is implemented as a standard part of the pipeline for the entire survey, it tends to be applied more often in the case of \bp\ spectra. 

The reduced and calibrated spectra are analyzed using Redrock \citep[Bailey et al., in prep]{Redrock_Anand2024}, a template-based Principal Component Analysis (PCA) method, to measure redshifts and determine spectral classifications (in three broad categories of {\tt STAR}, {\tt QSO}, or {\tt GALAXY}) for all sources. Although the vast majority of \bp\ sources are stars, the selection described in Table~\ref{tab:targetselection} does result in a small number of extragalactic sources (see \S~\ref{sec:science_yield} for more details). 

After the Redrock processing is complete, we run two additional pipelines (RVS and SP), which fit stellar atmosphere models to the spectra and compute more accurate radial velocities, stellar parameters (effective temperatures (\Teff), surface gravities (log\,$g$), metallicities, [$\alpha$/Fe], etc.), and a few elemental abundances for all the sources in the \bp. The sources are also crossmatched to the {\it Gaia} DR3 catalog. These quantities are included as part of the 
DESI Stellar Catalog, a Value-Added Catalog released as part of the DESI EDR \cite{Koposov2024_EDR_VAC} and DR1 \cite{DESI_DR1_2025,Koposov2025}. 

\subsection{Data Quality and Velocity Uncertainties}

\begin{figure*}
\centering
\includegraphics[width=\linewidth]{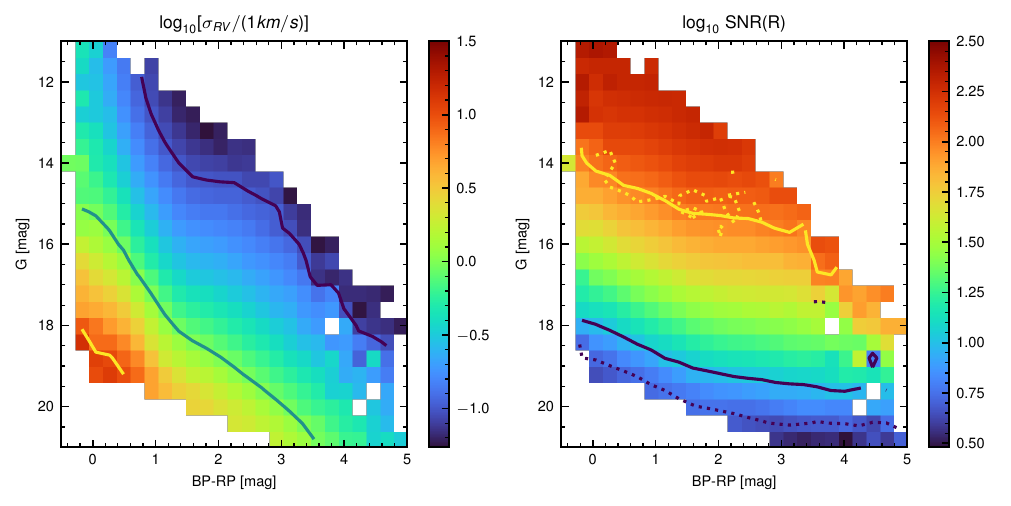}
\caption{(Left panel) Formal radial velocity uncertainties as a function of {\it Gaia} $G$ magnitude and $(BP-RP)$ color. The colorbar shows ${\rm log_{10}(\sigma_{RV}/[1~{\rm km\,s^{-1}}])},$ and lines show contours of constant radial velocity uncertainty of 0.1,1 and $10~{\rm km\,s^{-1}}$ (dark blue, light blue, and yellow respectively). (Right panel) Median signal-to-noise ratio per pixel in the R-arm of the spectrograph as a function of {\it Gaia} colour and magnitude. The solid lines show contours corresponding to SNR=10 (blue) and SNR=100 (yellow). The dotted lines show the same SNR contours but for the bright-time \mws.}
\label{fig:velocityprecision}
\end{figure*}

Figure~\ref{fig:velocityprecision} shows the distribution of formal radial velocity uncertainties as a function of the {\it Gaia} $G$ and $(BP-RP)$ colors as measured in the DESI DR2 data set. The bulk of the sample has velocity measurement uncertainties of $\le 5$~km/s, with only the bluest (i.e., $(BP-RP) \lesssim 1$) and faintest (i.e., $G \gtrsim 18$) populations having larger uncertainties (i.e., $\gtrsim 10$~km/s). In addition to the formal uncertainties, a comparison of repeat observations of sources show a systematic floor to the noise level of $\approx 2~{\rm km\,s^{-1}}$ for \bp\ spectra, higher than the noise floor of $\approx 1~{\rm km\,s^{-1}}$ observed for the \mws\ \citep[for a detailed discussion, please see][]{DESI_DR1_2025,Koposov2025}. 

In the process of analyzing the velocity accuracy of the data from the \bp\ included in the DESI DR1 release,  
a problem was identified with the wavelength calibration of some exposures.  Due to the increased continuum sky brightness and the 
algorithm used to correct the wavelength solution by using sky lines, incorrect offsets were applied, which resulted in some exposures having large (in rare cases up to $\sim 20~{\rm km\,s^{-1}}$) 
velocity zeropoint offsets. This algorithm has since been modified to be less affected by 
high sky brightness
and has been corrected for DESI DR2. However, the DESI DR1 radial velocities for all sources have a median offset of $\Delta V \equiv V_{DR2} - V_{DR1} \approx 1~{\rm km\,s^{-1}}$, with a scatter of $\sigma(\Delta V)\approx 2.6~{\rm km\,s^{-1}}$. In addition, the distribution of $\Delta V$ is asymmetric and has significant non-gaussian tails, with $\approx$8\% of sources having $\vert \Delta V \vert \ge 10~{\rm km\,s^{-1}}$ \citep[see][for details]{Koposov2025}.  When using the \bp\ data from DR1, we caution the reader to 
use the correction from \citet[]{Koposov2025} and assume additional systematic errors $<5-10~{\rm km\,s^{-1}}$ depending on the field location \citep[see][for details]{Koposov2025}.  


\subsection{Gaia RVS survey comparison}

\begin{figure}[hb]
\centering
\includegraphics[width=\linewidth]{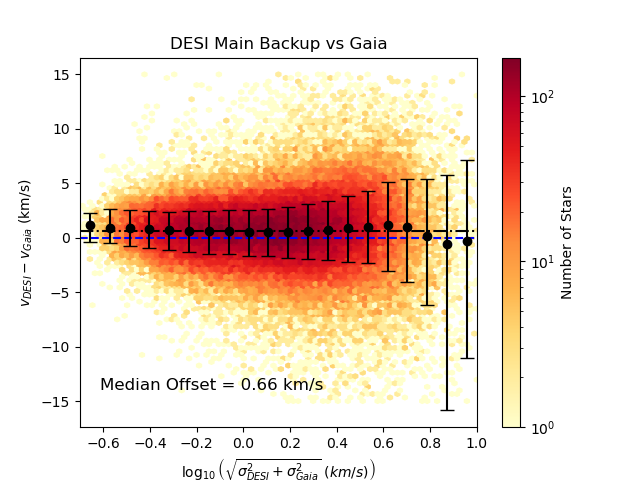}
\caption{Comparison between \bp\ and {\it Gaia} radial velocity measurements for $\approx 10^5$ stars from the DESI DR2 release with $>20\sigma$ {\it Gaia} RVS 
measurements. The blue dashed line shows $\delta(v)\equiv v_{DESI}-v_{Gaia} = 0$, and the black points show the median value of the offset in bins of total error. The vertical bars show the range between the 16th and 84th percentiles. The dashed black line shows the median value of $0.66~{\rm km\,s^{-1}}$, which represents the overall systematic offset between the DESI DR2 and {\it Gaia} radial velocity values.}
\label{fig:MWBP_Gaia_Comparison}
\end{figure}

The \bp\ radial velocities agree well with the velocities reported by the $Gaia$ Radial Velocity Spectrometer (RVS). $Gaia$ RVS  undertook high-resolution ($R\equiv\lambda/\delta\lambda \approx 11,500$) spectroscopy in a narrow wavelength range (8460 - 8700\AA) of roughly 5.6 million bright ($G\lesssim 14$~mag) stars \citep{Gaia_RVS_2023}. Approximately 25\% of all the sources observed by \bp\ overlap with these; in contrast, $<7\%$ of sources in the \mws\ 
have {\it Gaia} RVS measurements. For the sample that overlaps, DESI contributes spectral information spanning the entire 3600-9800\AA\ wavelength range. Figure~\ref{fig:MWBP_Gaia_Comparison} compares the \bp\ and {\it Gaia} radial velocities for stars from the DESI DR2 release with $\ge 20\sigma$ {\it Gaia} radial velocity measurements. These show a small residual systematic offset of $\delta(v)\equiv v_{DESI}-v_{Gaia}\approx0.7~{\rm km\,s^{-1}}$. 

\subsection{Scientific Yield}
\label{sec:science_yield}



As of 2025 March 1, DESI has completed roughly 70\% of the tiles for the \bp, resulting in spectra of $>$7 million stellar targets. DESI DR1
presents catalogs and spectra of $1.2$M of these. 
The DESI DR1 Stellar Catalog
presents radial velocities, effective temperatures 
\Teff\  surface gravities (log $g$), metallicities and a few elemental abundances of the stars derived using the RVS and SP 
pipelines \citep{Koposov2024_EDR_VAC,Koposov2025}.

\begin{figure}[ht]
    \centering
    \includegraphics[width=\linewidth]{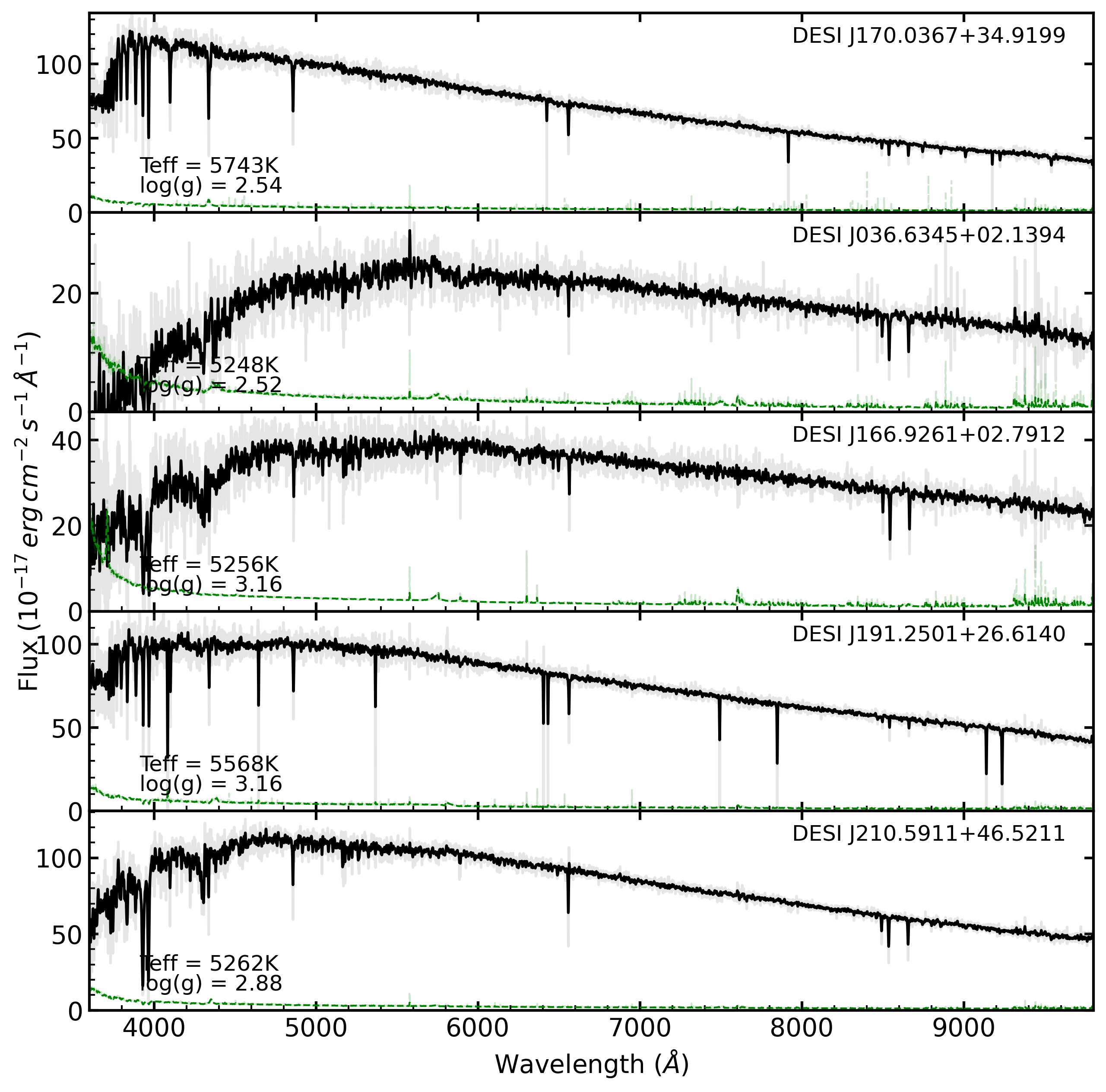}
    \caption{Spectra of five random giant stars of target class {\tt BACKUP\_GIANT} from the \bp. In each panel the light grey and green lines show the unsmoothed flux and noise spectra, and the black and darker green lines show the corresponding spectra smoothed by a 0.8\AA\ Gaussian kernel.}
    \label{fig:example_spectra_giants}
\end{figure}

\begin{figure}[ht]
    \centering
    \includegraphics[width=\linewidth]{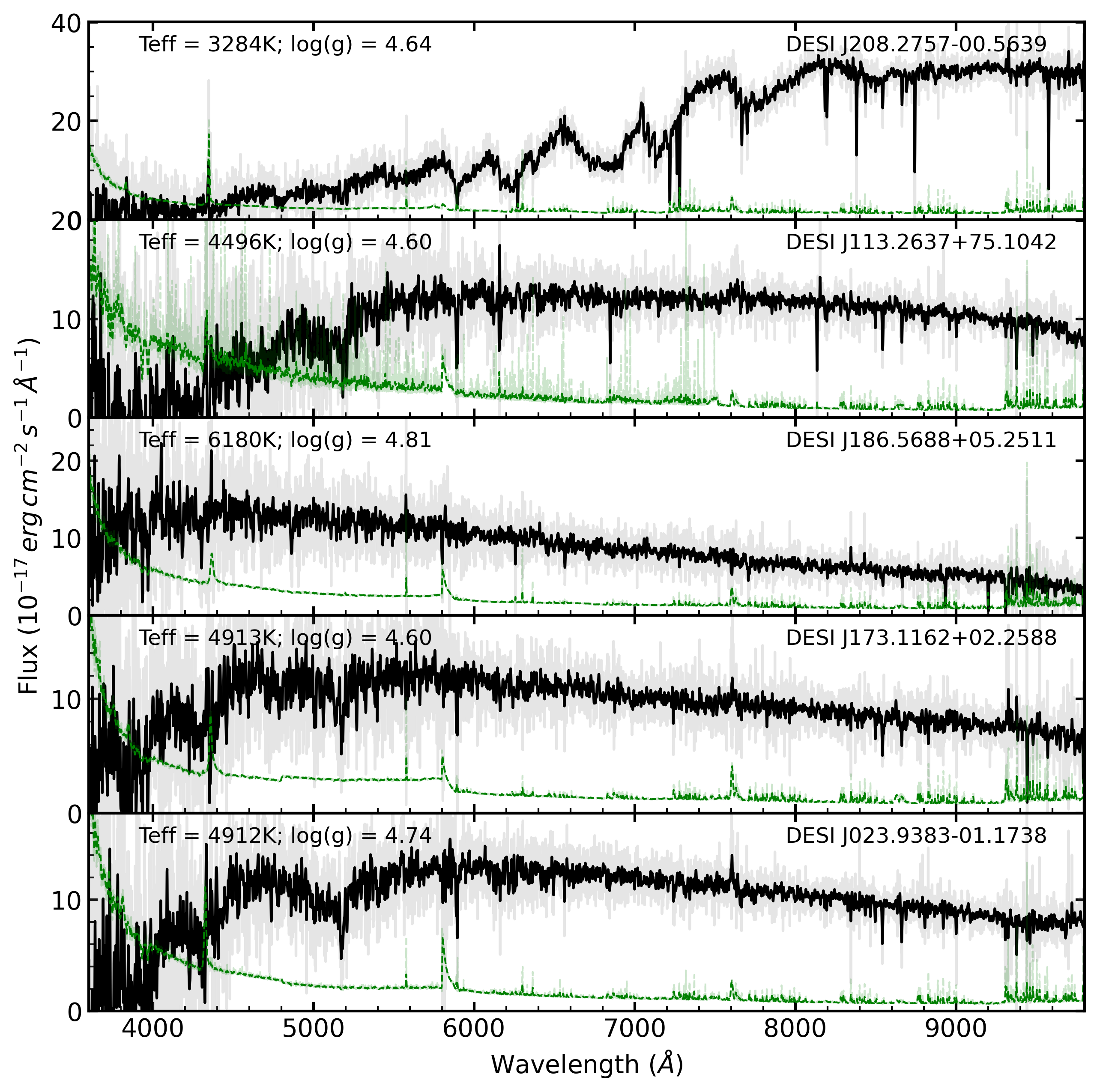}
    \caption{Spectra of five random stars of target class {\tt BACKUP\_VERY\_FAINT} from the \bp. In each panel the light grey and green lines show the unsmoothed flux and noise spectra, and the black and darker green lines show the corresponding spectra smoothed by a 0.8\AA\ Gaussian kernel.}
    \label{fig:example_spectra_vfaint}
\end{figure}

From the RVS results, we find the yield of giant stars (sources with ${\rm log} g  < 4$) is $\approx 21\%$ for the {\tt BACKUP\_GIANT} targets and $\approx$ 14\% for the {\tt BACKUP\_GIANT\_LOP} targets. Example spectra of these giant stars are shown in Figure~\ref{fig:example_spectra_giants}.  
In addition, about 15\% of the {\tt BACKUP\_BRIGHT} stars are classified as giants. Figure~\ref{fig:example_spectra_vfaint}
shows five example spectra of the {\tt BACKUP\_VERY\_FAINT}  target class to illustrate the data quality and kinds of sources. 

\begin{figure}[ht]
    \centering
    \includegraphics[width=\linewidth]{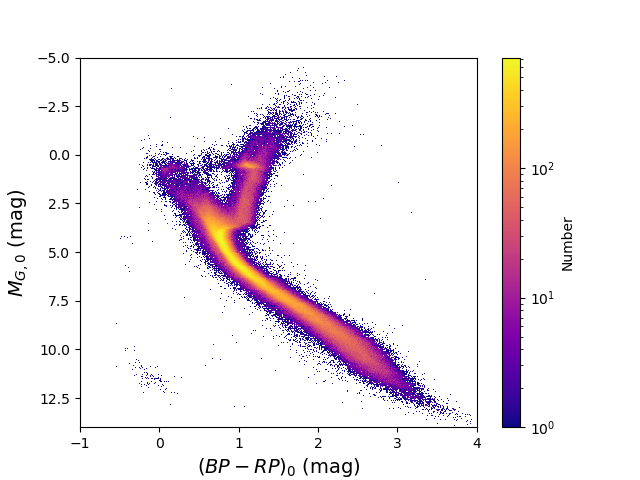}
    \caption{Extinction-corrected observational Hertzsprung-Russell diagram of stars from the \bp, constructed using distance estimates from the SpecDis Value-Added Catalog \citep{SpecDis_2025}.}
    \label{fig:hr_diagram}
\end{figure}

The Milky Way Survey SpecDis Value Added Catalog \citep[SpecDIS VAC;][]{SpecDis_2025} presents distance estimates for nearly all stars in the DESI~DR1 release (including MWBP targets) based on a neural network approach. Figure~\ref{fig:hr_diagram} shows the observational Hertzprung-Russell diagram 
in the {\it Gaia} absolute $G$-band magnitude vs.~$(BP-RP)$ color plane for the $\approx 0.76$ million \bp\ stars from DESI DR1 with distance estimate uncertainties $\lesssim 25\%$. The {\it Gaia} magnitudes and colors have been corrected for Galactic dust extinction using the coefficients derived by \citet{Gaia_DR2_HR_Diagram} and the $E(B-V)$ estimates derived by \citet{SFD1998} with corrections from \citet{Schlafly2011}. Figure~\ref{fig:hr_diagram} demonstrates that the \bp\ covers most of the stellar locus, including a small number of white dwarfs. 

\begin{figure*}[t]
    \centering
    \includegraphics[width=\linewidth]{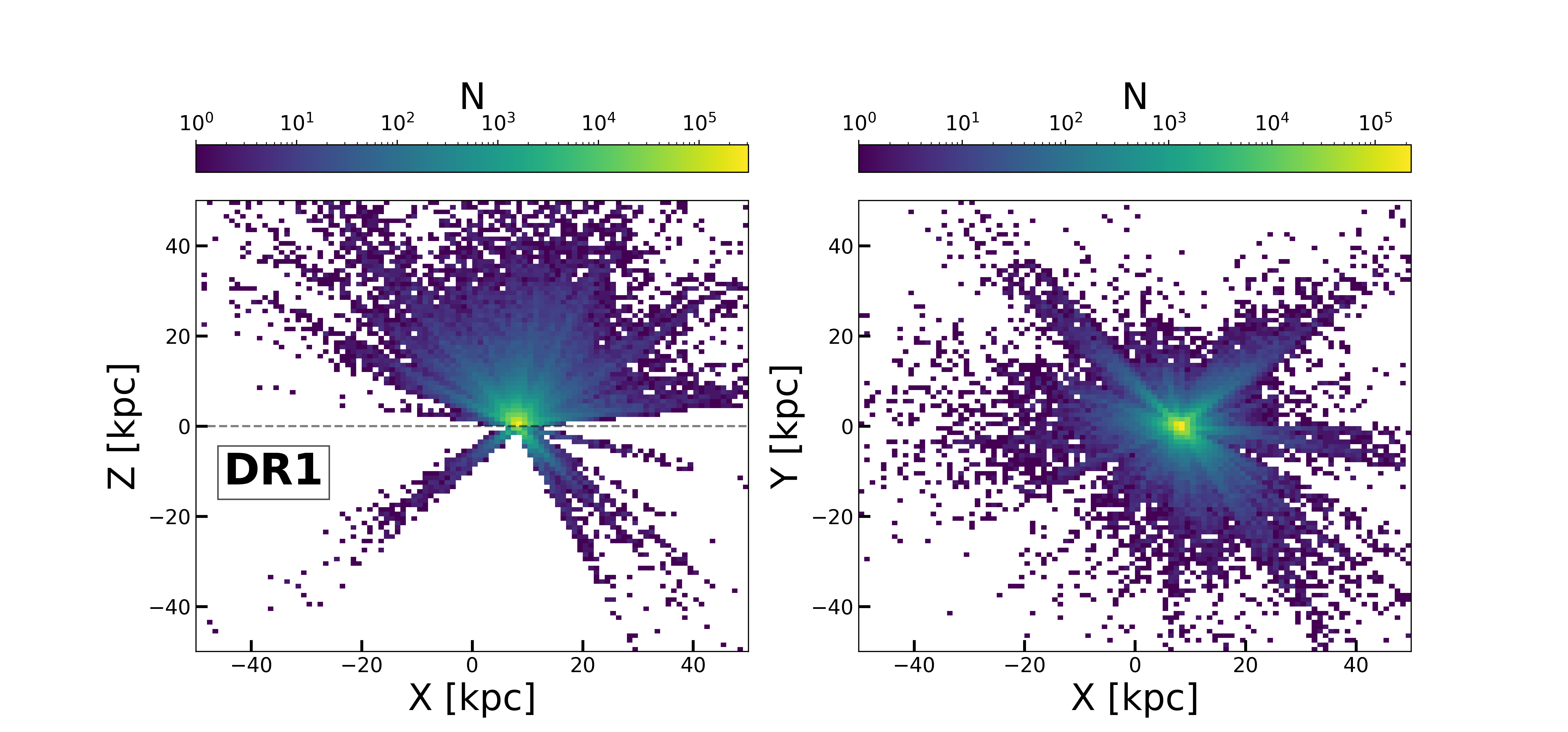}
    \includegraphics[width=\linewidth]{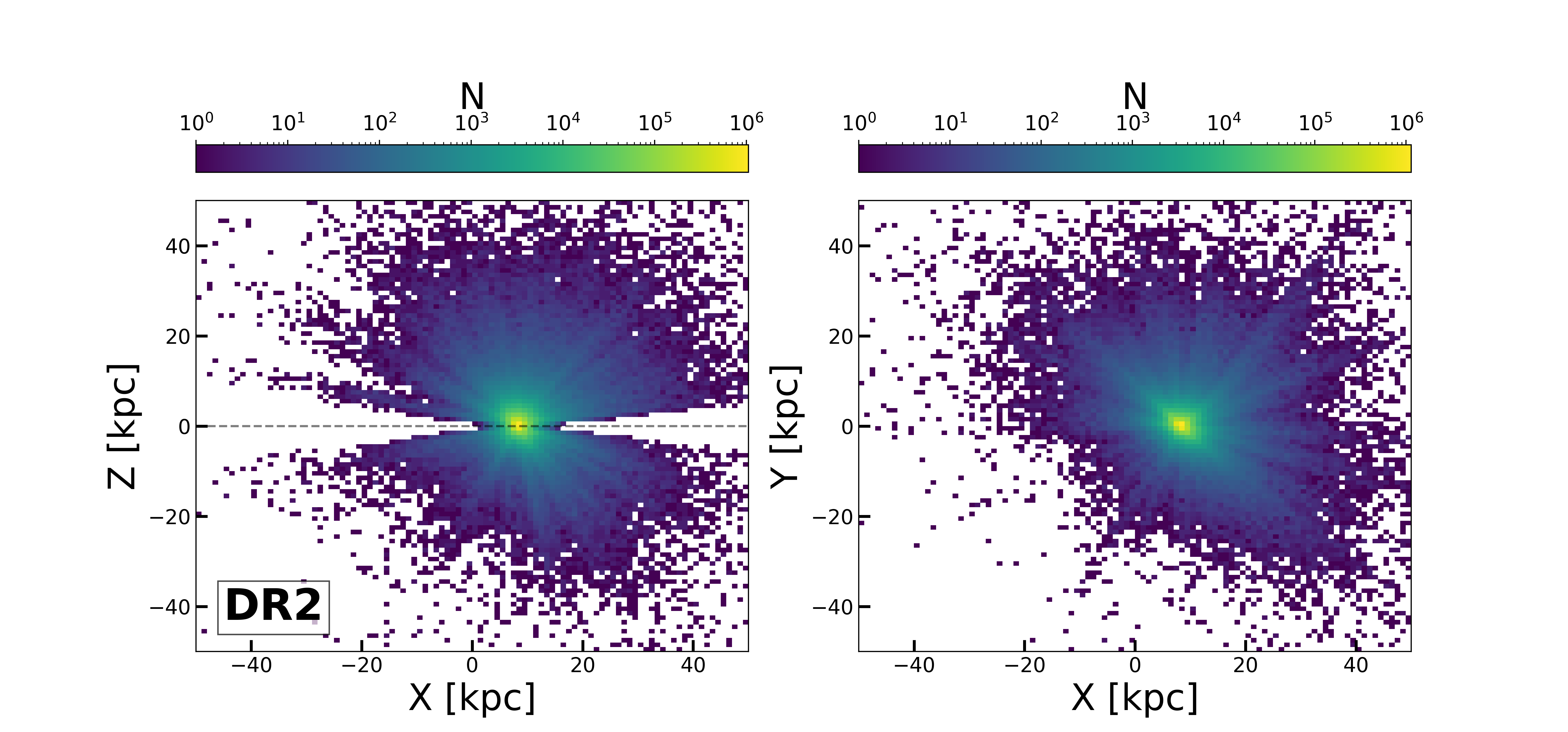}   
    \caption{The projection of spectroscopically observed MWBP targets in Galactocentric coordinates (X,Y,Z) of the \bp\ based on distances from the 
    SpecDis Catalog \citep[]{SpecDis_2025}, color-coded by number density of spectra. The top and bottom panels show the density distribution of stars for the DESI DR1 and DR2 data releases respectively, which represent $\approx12\%$ (DR1) and nearly 50\% (DR2) of the total \bp~footprint (cf.~left panel of Figure~\ref{fig:backup_dr1_dr2}).}
    \label{fig:galdist_backup}
\end{figure*}

\begin{figure*}[t]
    \centering
    \includegraphics[width=\linewidth]{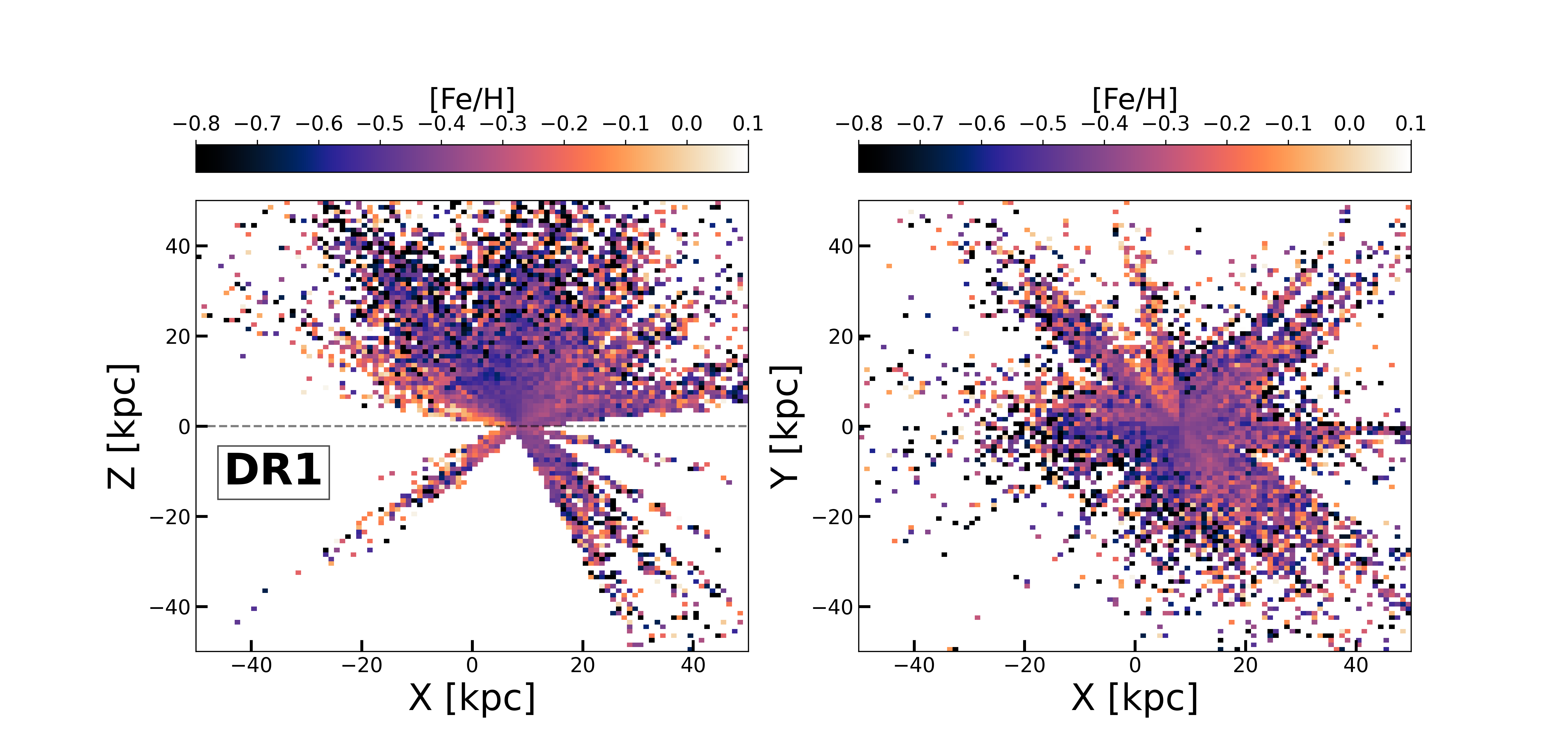}
    \includegraphics[width=\linewidth]{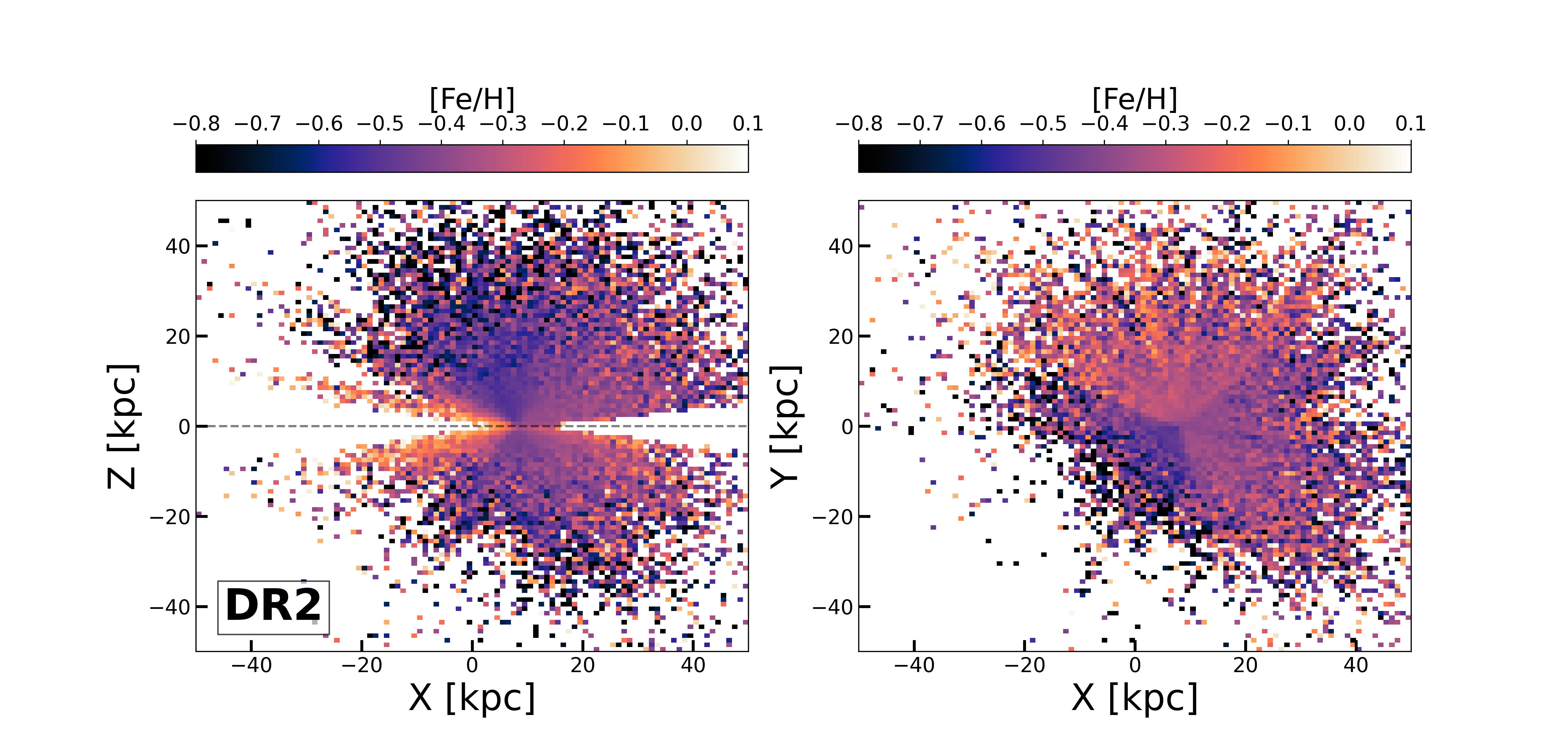}   
    \caption{The distribution of metallicities of spectroscopically observed \bp\ targets shown  in Galactocentric coordinates (X,Y,Z) of the \bp\ for the DESI DR1 and DR2 data releases respectively. Distance estimates are based on the SpecDis Catalog \citep[]{SpecDis_2025} and metallicities are from the RVS estimates \citep{Koposov2024_EDR_VAC,Koposov2025}. }
    \label{fig:galdist_backup_feh}
\end{figure*}

Figures~\ref{fig:galdist_backup} and \ref{fig:galdist_backup_feh} present the spatial distribution of the \bp\ DR1 and DR2 samples in Galactocentric coordinates, color-coded by number density and metallicity. The conversion to a Galactocentric frame assumed a Solar velocity of [11.1,248.5,7.25]~km~s$^{-1}$ \citep{Schonrich2010,Reid2020}, a Galactocentric distance of 8.178~kpc \citep{Gravity2019}, and an $Z$-distance offset of 20.8~pc \citep{Bennett_Bovy2019}. The distance and metallicity estimates are from the SpecDis VAC and DESI DR1 Stellar Catalog, respectively \citep[see, e.g.,][]{SpecDis_2025,Koposov2025}. 
The \bp\ stars are primarily located within 15 kpc of the Sun (top panels) and span a range in metallicity (bottom panels).

\subsubsection{Comparison with the PRISTINE Survey}

\begin{figure}[ht]
    \centering
    \includegraphics[width=\linewidth]{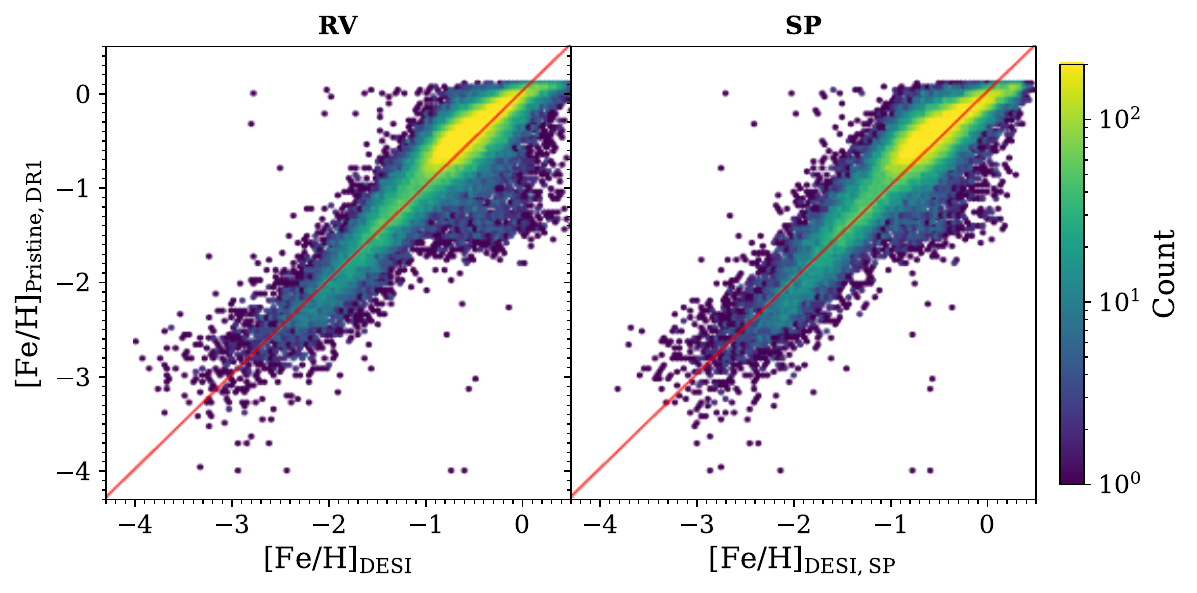}
    \caption{Comparison of the photometrically-derived metallicities in the PRISTINE DR1 catalog \citep{Martin2023} with the spectroscopic metallicity estimates from the \bp\ observations of the same stars based on the RVS (left) and SP (right) pipelines as reported in the DESI DR1 Stellar Catalog \citep{DESI_DR1_2025,Koposov2024_EDR_VAC,Koposov2025}. The solid red line in both panels is a 1-1 line, not a fit.}
    \label{fig:pristine_dr1_comparison}
\end{figure}

The \bp\ metallicities 
can also be used to validate photometric metallicities derived from narrow-band imaging surveys. 
For example, the \bp\ 
provides spectroscopic measurements for a significant population of stars measured by the PRISTINE Survey \citep{Starkenburg2017,Martin2023}, which uses
MegaCam on the Canada-France-Hawaii Telescope to map the northern sky in a narrow-band CaHK filter during poor weather time, 
an effort not unlike \bp. 
Data Release 1 of the PRISTINE Survey reports photometric and astrometric measurements for approximately 6.38M stars over 6500~deg$^2$, along with photometrically estimated metallicity measurements \citep[see][]{Martin2023}. The DESI DR1 and DR2 data overlap with this sample over $\approx$925~deg$^2$ and $\approx$3,300~deg$^2$, respectively.  Figure~\ref{fig:pristine_dr1_comparison} compares the photometric metallicities in the PRISTINE DR1 catalog with the (more accurate) spectroscopic metallicities from the DESI 
DR1 Stellar Catalog 
\citep{DESI_DR1_2025,Koposov2025} for the stars meeting the quality criteria listed in \citet{Martin2023} and the quality criteria related to the SP pipeline described in \citet{Koposov2024_EDR_VAC}.
We confirm that the PRISTINE photometric metallicity estimates are generally quite good, with a scatter of 0.22 (0.23)~dex compared to the 
SP (RVS) pipeline spectroscopic metallicities. However, PRISTINE metallicities are systematically more metal poor by 0.06 (0.05)~dex compared to the DESI 
SP (RVS) metallicities, respectively. This offset decreases to 0.02 (0.01)~dex below [Fe/H]$<-2.0$. 

While these differences are small, we caution the reader that they are also much smaller than the accuracy of the DESI spectroscopic metallicities. \citet{Koposov2025} cross matches the stars in the DESI DR1 Stellar Catalog with those measured by the APOGEE, GALAH, Gaia and SAGA surveys and finds that the DESI metallicity estimates show systematic offsets relative to the other surveys, which are different for giant and dwarf stars. Empirically correcting for these systematic offsets results in scatter of the DESI metallicity estimates of $\approx0.1$~dex.

\subsubsection{Extragalactic Sources in \bp}

In addition to obtaining spectra of a large number of stars, the \bp\ also obtains spectra of extragalactic sources that are included in the {\it Gaia} catalog. 
Of the 1.2M sources in the DESI DR1 release with reported reliable Redrock measurements (i.e., {\tt ZWARN}=0 and {\tt DELTACHI2} $\ge$ 25), Redrock classifies 2743 ($\approx 0.23\%$) as spectral type {\tt QSO} and 16,240 ($\approx1.34\%$) as type {\tt GALAXY}. The set classified by Redrock as spectral type {\tt QSO} contains only a small number of misclassified sources (a few Carbon stars, unusual white dwarfs, and a handful of sources with calibration issues). 
Such QSO misclassifications are a minority: only 27 ($<0.1\%$) 
of sources classified as {\tt QSO} by Redrock
are Galactic stars. 

The set classified by Redrock as {\tt GALAXY}, in contrast, includes a large fraction ($\approx 65\%$) of misclassified stars. These include reddened stellar sources, a handful of sources that show
emission lines from interstellar or circumstellar gas, or stellar spectra that are poorly represented by the limited set of Redrock templates. 

\begin{figure}[ht]
    \centering
    \includegraphics[width=\linewidth]{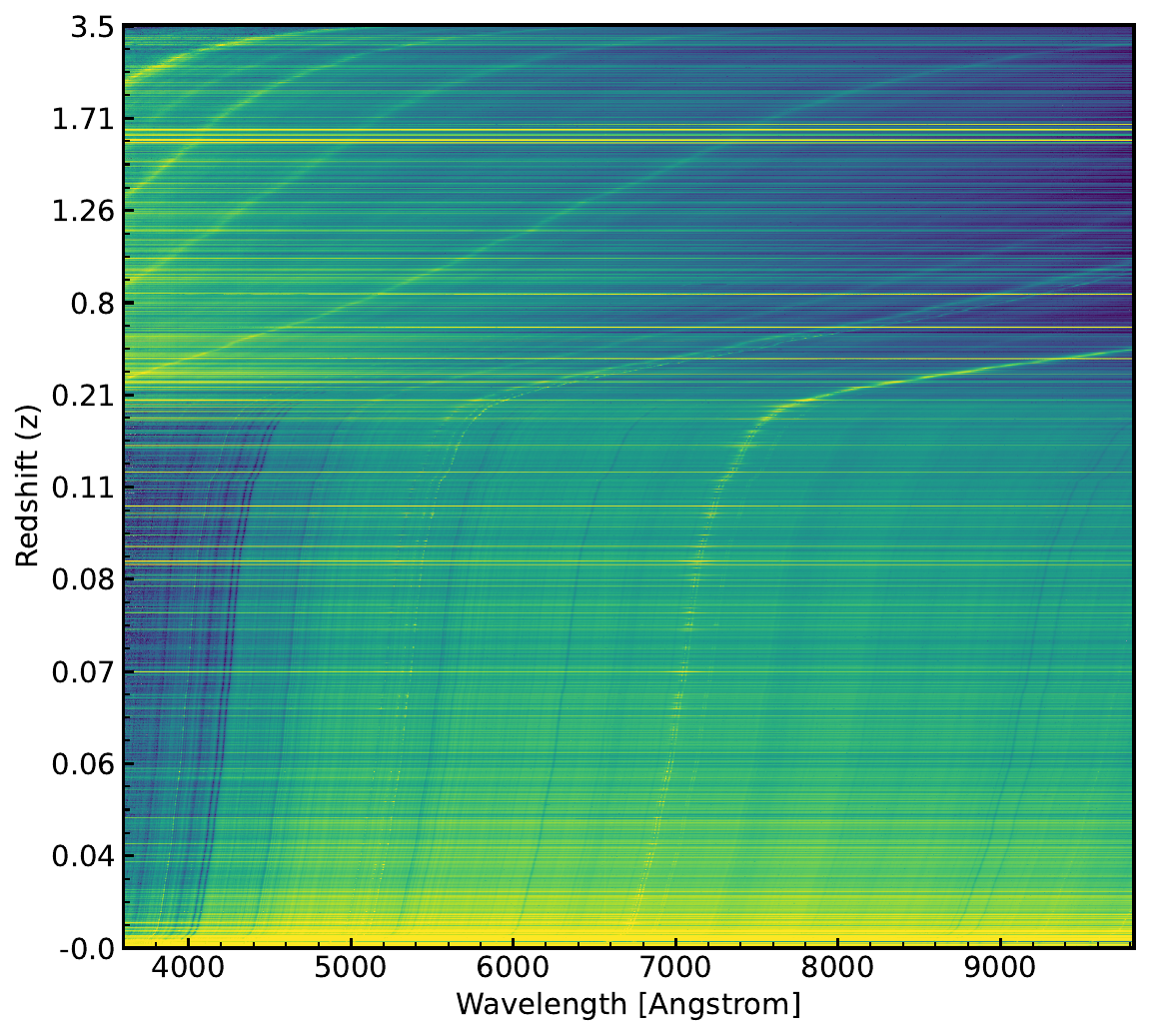}
    \caption{5687 DESI spectra of primarily extragalactic sources from the \bp, arranged 
   in increasing redshift. For all of these spectra, the DESI Redrock pipeline produced $\ge5\sigma$ redshift determinations, but the RVS pipeline failed to find a fit. 
   Several emission and absorption features typical of galaxy and QSO spectra are seen tracking diagonally across the image with increasing redshift. A small number of the brightest horizontal lines are spectra of stars that are misclassified as spectral type {\tt GALAXY} by Redrock.}
    \label{fig:exgal_vs_z}
\end{figure}

\begin{figure}[ht]
    \centering
    \includegraphics[width=\linewidth]{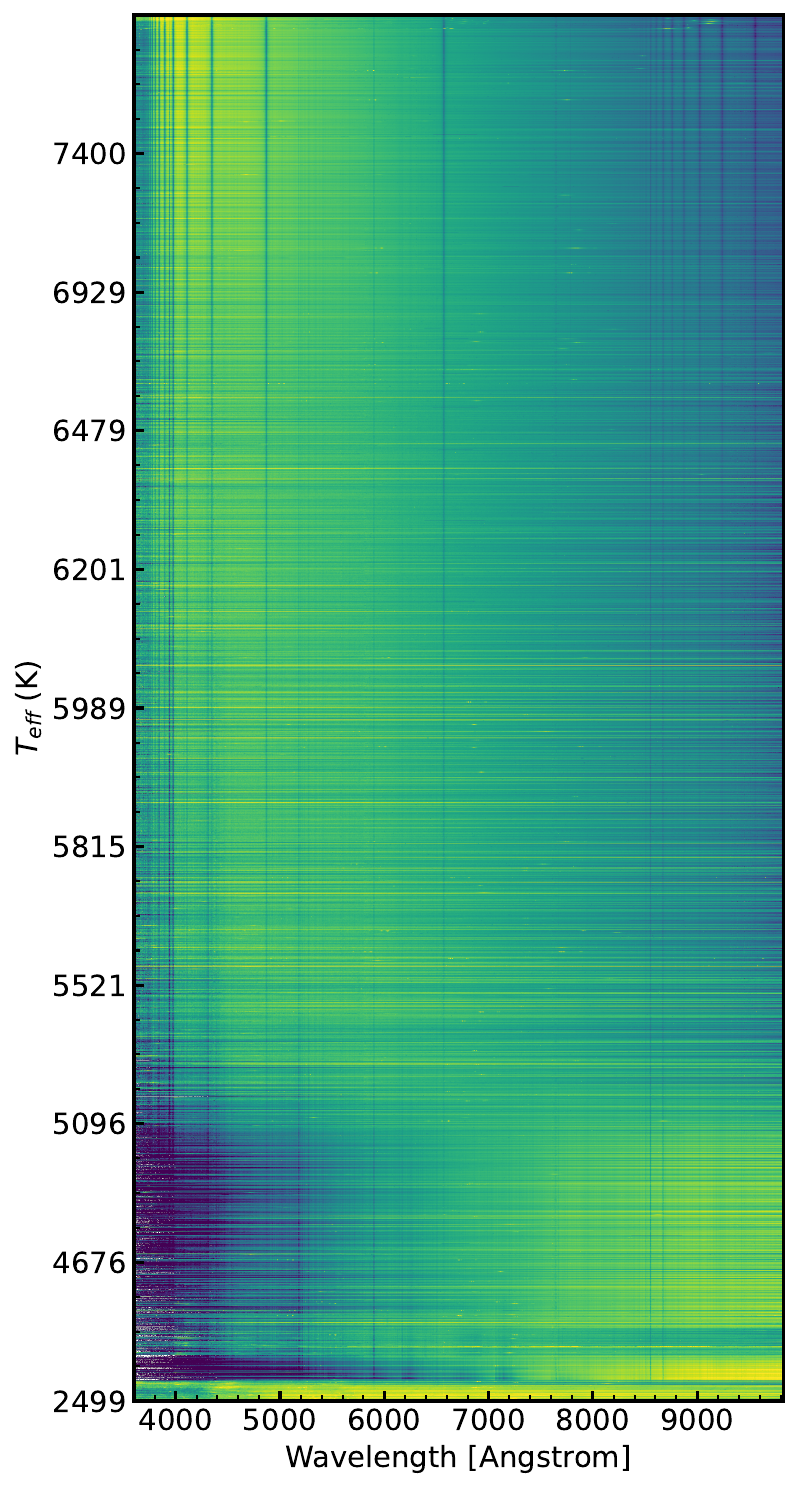}
    \caption{Spectra of 13,290 (primarily) stellar sources from the \bp\ that were misclassified as spectral type {\tt GALAXY} by the Redrock pipeline, but for which the RVS pipeline found a good model fit. The sources are sorted by their effective temperature estimated by the RVS pipeline. Nearly all of the sources are stars, as evidenced by their (approximately) zero redshift absorption features. However, there are also several galaxies and QSOs visible within this set, which are identifiable by their emission signatures. The sources with the lowest estimated effective temperatures (\Teff\ $<3000$K) are also likely sources with problematic RVS fits.}
    \label{fig:stars_misclassified}
\end{figure}

Galaxy and QSO spectra from the \bp\ can generally be selected from the DR1 Stellar Catalog by identifying the sources that 
Redrock has confidently identified as a {\tt GALAXY} or {\tt QSO} spectral type (i.e., {\tt ZWARN=0} and {\tt DELTACHI2}$\ge$25), and for which 
the RVS fit results in warning flags 
(i.e., {\tt RVS\_WARN}$\ne$0). Figure~\ref{fig:exgal_vs_z} shows these spectra ordered by redshift, 
where the pattern of redshifting emission lines typical of galaxies (at $z<1.6$) and quasars (at $z>2$) is readily seen. 
This plot also shows some outliers (e.g., the two bright continuum spectra which can be seen near $z\approx 1.7$) which tend to be the stellar contaminants. Figure~\ref{fig:stars_misclassified} shows the sources that 
Redrock robustly identified as galaxies or QSOs, but for which the RVS model fits had no warning flags (i.e., {\tt RVS\_WARN}=0). Nearly all of these sources are indeed stars. However, close examination of Figure~\ref{fig:stars_misclassified} reveals a small but significant number of galaxy and QSO emission lines scattered within the collection (e.g., several spectra which show emission lines at $\sim7800$\AA\ in the upper right quadrant of the plot).%

\begin{figure}[ht]
    \centering
    \includegraphics[width=\linewidth]{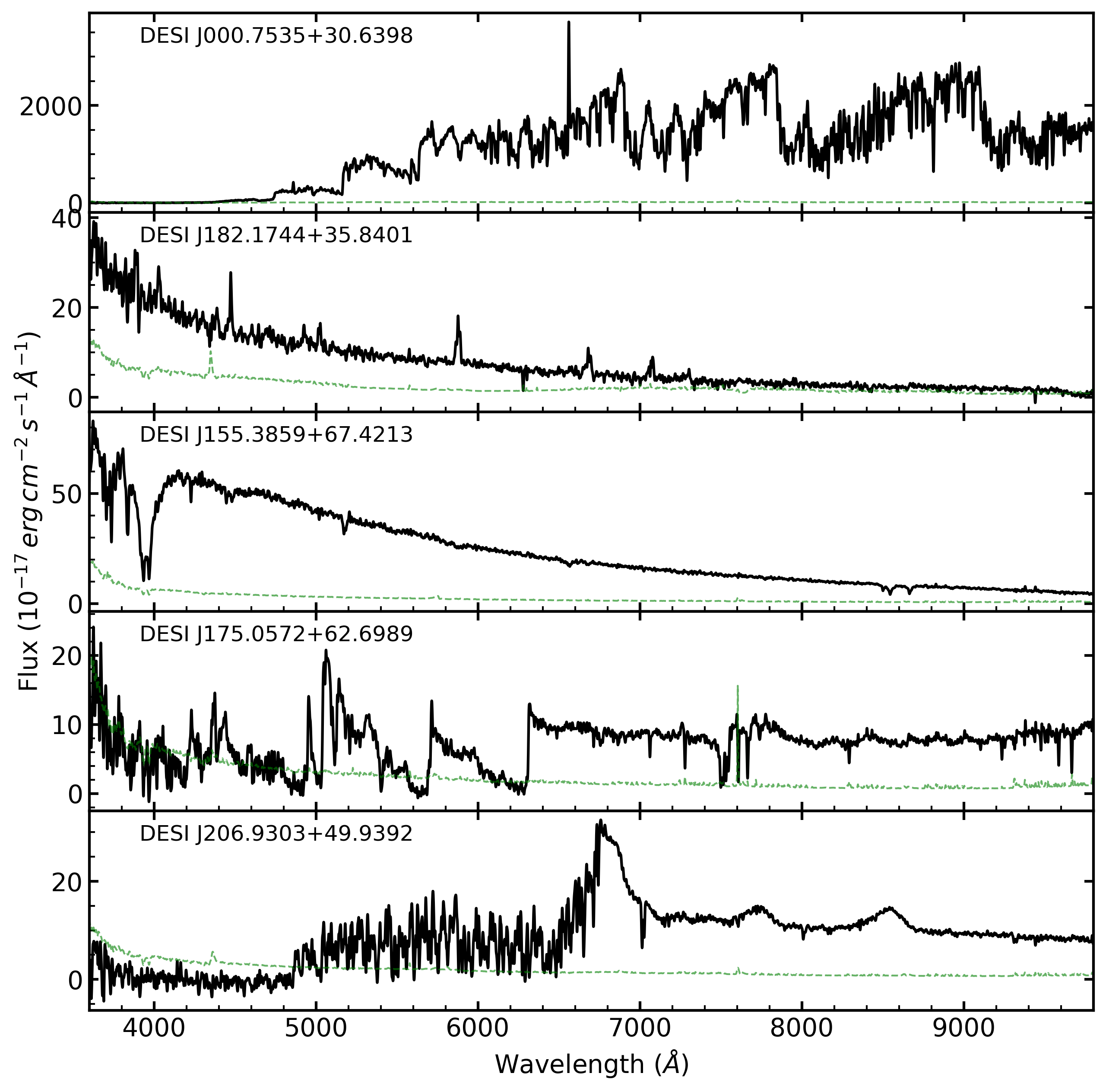}
    \caption{Smoothed spectra of a few rare gems from the \bp. From top to bottom, the panels show 
    a Carbon Star, 
    a AM CVn type cataclysmic variable, 
    an metal enriched white dwarf with strong Ca lines, 
    a broad absorption line QSO at $z=3.077$ and 
    a high-redshift QSO at $z=4.526$. The green lines show the noise level in these spectra.}
    \label{fig:example_spectra_gems}
\end{figure}

Figure~\ref{fig:example_spectra_gems} shows examples of five rare objects observed by the \bp, illustrating the range of astronomical gems produced by large spectroscopic surveys. The bottom two panels of Figure~\ref{fig:example_spectra_gems} show examples of QSO spectra from the \bp; the top three panels show sources that were misclassified as type {\tt QSO} by Redrock, but which turn out to be rare stellar sources. 

\section{Summary}
\label{sec:summary}

We have presented an overview of the Milky Way Backup Program (\bp) being carried out by the Dark Energy Spectroscopic Instrument at the Mayall 4-m Telescope at the Kitt Peak National Observatory. The goal of the \bp\ is to utilize 
twilight times ($<18^\circ$) and poor weather conditions as efficiently as possible 
to supplement the \mws. 
Accordingly, the \bp\ extends the observed range of targets to brighter magnitudes and lower Galactic latitude and declination than that of the \mws.
The final \bp\ will cover more than 20,000~deg$^2$ at declinations $> -28.5^\circ$ and $\vert b \vert \gtrsim 7^\circ$ and deliver spectroscopy of roughly ten million stars. 

The \bp\ targets are selected from the {\it Gaia} catalog, prioritizing an unbiased sample of bright stars ($11.2\le G <16$) and candidate halo giant stars ($16< G \lesssim 19$) selected based on their {\it Gaia} (DR2) parallaxes. The \bp\ has the advantage of obtaining multiple (low signal-to-noise ratio) observations for many of the  targets with nearly 60\% of the \bp\ stars having more than one epoch of observation. While the data are of poorer quality than the \mws\
for stars of similar magnitudes, the \bp\ results in spectra with average $R$-arm signal-to-noise ratio exceeding 100 for the brightest stars, and delivers radial velocity measurements with a precision of $\approx 1~{\rm km\,s^{-1}}$ for stars with {\it Gaia} $G\approx 18-19$~mag (depending on color). 

The DESI DR1 release, which represents (roughly) the first year of DESI survey observations, includes spectra of $\approx$1.2M unique \bp\ targets, and covers $\approx$12\% of the total footprint. 
The stellar measurements for the \bp\ based on DR1 were released as part of the DESI DR1 Stellar Catalog and are described in \citet{Koposov2025}. 
The survey has currently completed nearly 70\% of its originally planned footprint, and is likely to complete the original set of \bp\ tiles within a year. Given the success of this program, we will likely continue it in future phases of the DESI project by increasing the density of the tiling and / or increasing the areal coverage as appopriate. The \bp\ represents a good example of how to optimize the utility of poor weather time for ongoing and future spectroscopic surveys.

\section*{Acknowledgements}

This material is based upon work supported by the U.S. Department of Energy (DOE), Office of Science, Office of High-Energy Physics, under Contract No. DE–AC02–05CH11231, and by the National Energy Research Scientific Computing Center, a DOE Office of Science User Facility under the same contract. Additional support for DESI was provided by the U.S. National Science Foundation (NSF), Division of Astronomical Sciences under Contract No. AST-0950945 to the NSF’s National Optical-Infrared Astronomy Research Laboratory; the Science and Technology Facilities Council of the United Kingdom; the Gordon and Betty Moore Foundation; the Heising-Simons Foundation; the French Alternative Energies and Atomic Energy Commission (CEA); the National Council of Humanities, Science and Technology of Mexico (CONAHCYT); the Ministry of Science, Innovation and Universities of Spain (MICIU/AEI/10.13039/501100011033), and by the DESI Member Institutions: \url{https://www.desi.lbl.gov/collaborating-institutions}. Any opinions, findings, and conclusions or recommendations expressed in this material are those of the author(s) and do not necessarily reflect the views of the U. S. National Science Foundation, the U. S. Department of Energy, or any of the listed funding agencies.
The authors are honored to be permitted to conduct scientific research on I'oligam Du'ag (Kitt Peak), a mountain with particular significance to the Tohono O’odham Nation.


This work has made use of data from the European Space Agency (ESA) mission
{\it Gaia} (\url{https://www.cosmos.esa.int/gaia}), processed by the {\it Gaia}
Data Processing and Analysis Consortium (DPAC,
\url{https://www.cosmos.esa.int/web/gaia/dpac/consortium}). Funding for the DPAC
has been provided by national institutions, in particular the institutions
participating in the {\it Gaia} Multilateral Agreement.

This paper was begun at a workshop hosted by the Dunlap Institute for Astronomy and Astrophysics at the University of Toronto. We gratefully acknowledge the Dunlap Institute for their generous support and hospitality during the event, as well as Dunlap's Conferences \& Workshop Funding that helped make this workshop possible. The Dunlap Institute is funded through an endowment established by the David Dunlap family and the University of Toronto.

APC is supported by Taiwan's National Science and Technology Council (113-2112-M-007-009) and a Yushan Fellowship awarded by the Taiwanese Ministry of Education (MOE-113-YSFMS-0002-001-P2). SK acknowledges support from the Science \& Technology Facilities Council (STFC) grant ST/Y001001/1.

ADM was supported by the U.S.\ Department of Energy, Office of Science, Office of High Energy Physics, under Award Number DE-SC0019022

For the purpose of open access, the author has applied a Creative
Commons Attribution (CC BY) licence to any Author Accepted Manuscript version
arising from this submission.

\facility{Mayall (DESI),Gaia}
\software{Astropy \citep{astropy-collaboration13a, astropy-collaboration18a, astropy-collaboration22a}, 
fitsio (\url{https://github.com/esheldon/fitsio}),
healpy \citep{zonca19a},
Matplotlib \citep{hunter07a},
NumPy \citep{harris20a},
SciPy \citep{virtanen20a},
SPARCL \citep{juneau24a}.
}





\appendix

\begin{figure}[ht]
    \centering
    \includegraphics[width=\linewidth]{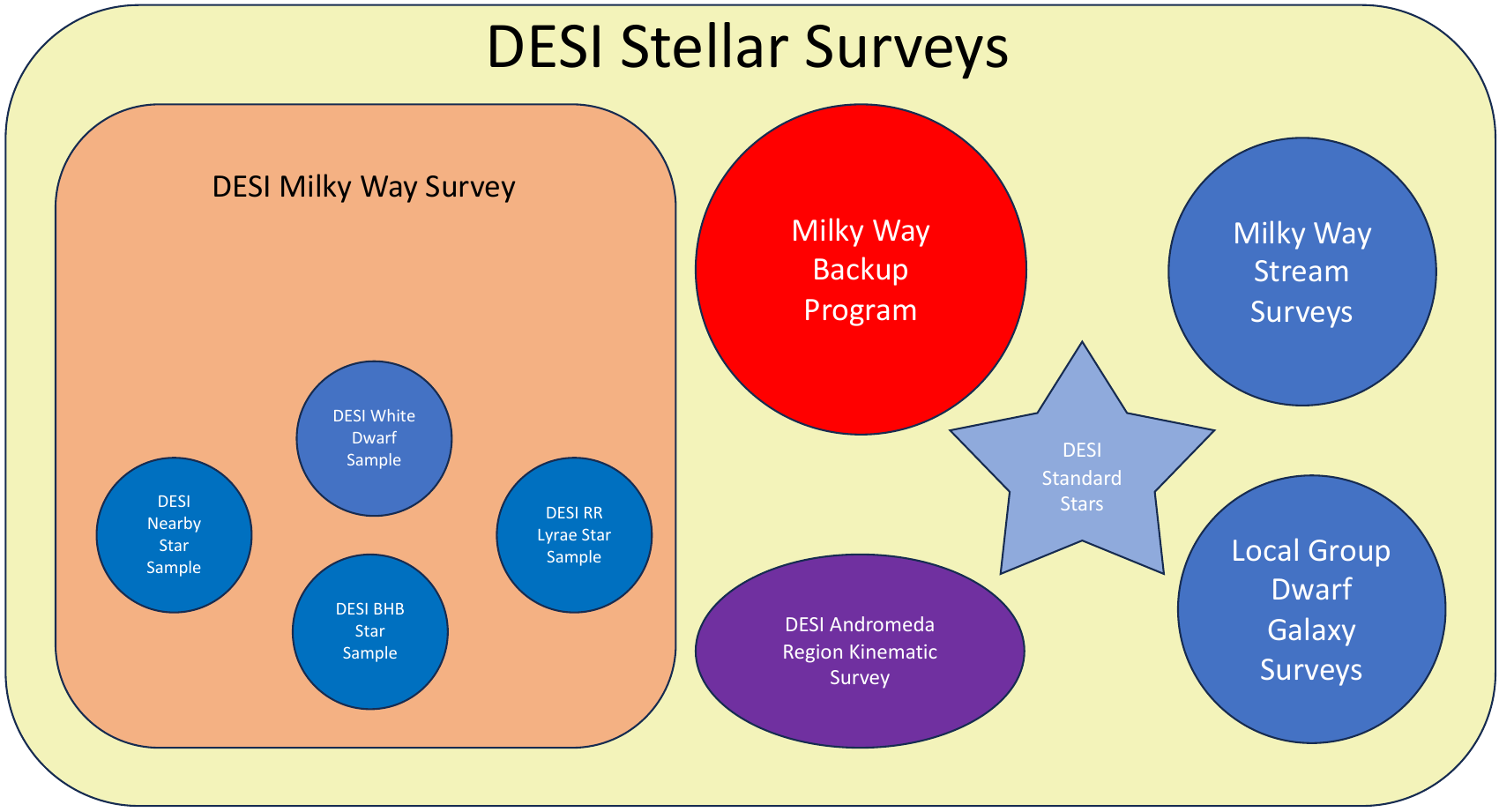}
    \caption{The current compendium and structure of the DESI Stellar Surveys. See text for details.}
    \label{fig:DESI_Stellar_Surveys}
\end{figure}

The nomenclature associated with the different DESI Stellar Surveys is presented in this Appendix and illustrated in Figure~\ref{fig:DESI_Stellar_Surveys}. The DESI Stellar Surveys consist of: (1) the Milky Way Survey (MWS) which is part of the DESI Main bright-time survey (and includes special samples of Nearby Stars, White Dwarfs, RR Lyrae and Blue Horizontal Branch stars), discussed in detail in \cite{Cooper2023}; (2) the Milky Way Backup Program (MWBP), which is the subject of this paper; (3) the DESI Andromeda Region Kinematic (DARK) Survey \citep[e.g.,][]{Dey2023_M31}; (4) standard stars observed by DESI on all dark, bright and backup tiles; (5) the Milky Way Stream Surveys \citep[e.g.,][]{Valluri2025_GD1}; and (6) the Local Group Dwarf Galaxy Surveys.  

The DESI Stellar Catalogs released thus far  \citep{Koposov2024_EDR_VAC,Koposov2025} include stellar sources observed by {\underline{\it all}} of these surveys and programs, and also include any other stars serendipitously targeted by other DESI observations (e.g., as part of the DESI Main dark-time survey, or by special DESI tiles).  
MWS and MWBP make up the bulk of the stars in the DESI Stellar Catalog. The DARK, Milky Way Stream and Local Group Dwarf Galaxy Surveys are targeted DESI observations made with special tiles, and (so far) constitute a small fraction of the Stellar Catalog. 

\bibliography{main}{}
\bibliographystyle{aasjournal}



\end{document}

%% file: author_list.tex
\author[0000-0002-4928-4003]{Arjun~Dey}
\affiliation{NSF NOIRLab, 950 N. Cherry Avenue, Tucson, AZ 85719, USA}

\author[0000-0003-2644-135X]{Sergey E. Koposov}
\affiliation{Institute for Astronomy, University of Edinburgh, Royal Observatory, Blackford Hill, Edinburgh EH9 3HJ, UK}
\affiliation{Institute of Astronomy, University of Cambridge, Madingley Rd, Cambridge CB3 0HA, UK}

\author[0000-0002-5758-150X]{Joan~R.~Najita}
\affiliation{NSF NOIRLab, 950 N. Cherry Avenue, Tucson, AZ 85719, USA}

\author[0000-0001-8274-158X]{Andrew P. Cooper}
\affiliation{Institute of Astronomy and Department of Physics, National Tsing Hua University, Hsinchu 30013, Taiwan}
\affiliation{Center for Informatics and Computation in Astronomy, National Tsing Hua University, Hsinchu 30013, Taiwan}

\author[0000-0002-2761-3005]{B.~T.~G\"ansicke}
\affiliation{Department of Physics, University of Warwick, Coventry, CV4 7AL, UK}

\author{Adam D.\ Myers}
\affil{Department of Physics and Astronomy, University of Wyoming, Laramie, WY 82071, USA}

\author[0000-0001-5999-7923]{A.~Raichoor}
\affiliation{Lawrence Berkeley National Laboratory, 1 Cyclotron Road, Berkeley, CA 94720, USA}

\author[0000-0002-2929-3121]{Daniel J.\ Eisenstein}
\affiliation{Center for Astrophysics $|$ Harvard \& Smithsonian, 60 Garden St., Cambridge MA 02138 USA}

\author[0000-0002-3569-7421]{E.~F.~Schlafly}
\affiliation{Space Telescope Science Institute, 3700 San Martin Drive, Baltimore, MD 21218, USA}

\author{C.~Allende~Prieto}
\affiliation{Instituto de Astrof\'{i}sica de Canarias, C/ Vía L\'{a}ctea, s/n, 38205 San Crist\'{o}bal de La Laguna, Santa Cruz de Tenerife, Spain}

\author[0000-0002-0740-1507]{Leandro {Beraldo e Silva}}
\affiliation{Steward Observatory and Department of Astronomy,\\University of Arizona, 933 N. Cherry Ave., Tucson, AZ 85721, USA}
\affiliation{Observatório Nacional, Rio de Janeiro - RJ, 20921-400, Brasil}

\author[0000-0002-9110-6163]{Ting S. Li}
\affiliation{Department of Astronomy and Astrophysics, University of Toronto, 50 St. George Street, Toronto ON, M5S 3H4, Canada}

\author[0000-0002-6257-2341]{M.~Valluri}
\affiliation{Department of Astronomy, University of Michigan, Ann Arbor, MI 48109, USA}

\author[0000-0002-0000-2394]{St\'ephanie~Juneau}
\affiliation{NSF NOIRLab, 950 N. Cherry Avenue, Tucson, AZ 85719, USA}

\author[0000-0002-2527-8899]{Mika Lambert}
\affiliation{Department of Astronomy \& Astrophysics, University of California, Santa Cruz, 1156 High Street, Santa Cruz, CA 95064, USA}

\author[0000-0002-6469-8263]{S.~Li}
\affiliation{Department of Astronomy, School of Physics and Astronomy, Shanghai Jiao Tong University, Shanghai 200240, China}
\affiliation{Shanghai Key Laboratory for Particle Physics and Cosmology, Shanghai 200240, China}
\affiliation{State Key Laboratory of Dark Matter Physics, School of Physics and Astronomy,Shanghai Jiao Tong University, Shanghai 200240, China}

\author[0000-0002-2468-5521]{Guillaume~F.~Thomas}
\affiliation{Instituto de Astrof\'isica de Canarias, E-38205 La Laguna, Tenerife, Spain \and Universidad de La Laguna, Dpto. Astrofísica, E-38206 La Laguna, Tenerife, Spain}

\author[0000-0002-5762-7571]{Wenting~Wang}
\affiliation{Department of Astronomy, School of Physics and Astronomy, and Shanghai Key Laboratory for Particle Physics and Cosmology, Shanghai Jiao Tong University, Shanghai 200240, People's Republic of China}

\author[0000-0001-5805-5766]{Alexander H.~Riley}
\affiliation{Institute for Computational Cosmology, Department of Physics, Durham University, South Road, Durham DH1 3LE, UK}
 
\author[0000-0003-0853-8887]{N.~Kizhuprakkat}
\affiliation{Institute of Astronomy and Department of Physics, National Tsing Hua University, Hsinchu 30013, Taiwan, Center for Informatics and Computation in Astronomy, National Tsing Hua University, Hsinchu 30013, Taiwan}


\author{J.~Aguilar}
\affiliation{Lawrence Berkeley National Laboratory, 1 Cyclotron Road, Berkeley, CA 94720, USA}

\author[0000-0001-6098-7247]{S.~Ahlen}
\affiliation{Department of Physics, Boston University, 590 Commonwealth Avenue, Boston, MA 02215 USA}

\author[0000-0003-4162-6619]{S.~Bailey}
\affiliation{Lawrence Berkeley National Laboratory, 1 Cyclotron Road, Berkeley, CA 94720, USA}

\author[0000-0001-9712-0006]{D.~Bianchi}
\affiliation{Dipartimento di Fisica ``Aldo Pontremoli'', Universit\`a degli Studi di Milano, Via Celoria 16, I-20133 Milano, Italy}
\affiliation{INAF-Osservatorio Astronomico di Brera, Via Brera 28, 20122 Milano, Italy}

\author{D.~Brooks}
\affiliation{Department of Physics \& Astronomy, University College London, Gower Street, London, WC1E 6BT, UK}

\author{T.~Claybaugh}
\affiliation{Lawrence Berkeley National Laboratory, 1 Cyclotron Road, Berkeley, CA 94720, USA}

\author[0000-0002-2169-0595]{A.~Cuceu}
\affiliation{Lawrence Berkeley National Laboratory, 1 Cyclotron Road, Berkeley, CA 94720, USA}

\author[0000-0002-1769-1640]{A.~de la Macorra}
\affiliation{Instituto de F\'{\i}sica, Universidad Nacional Aut\'{o}noma de M\'{e}xico,  Circuito de la Investigaci\'{o}n Cient\'{\i}fica, Ciudad Universitaria, Cd. de M\'{e}xico  C.~P.~04510,  M\'{e}xico}

\author[0000-0003-0928-2000]{J.~Della~Costa}
\affiliation{Department of Astronomy, San Diego State University, 5500 Campanile Drive, San Diego, CA 92182, USA}
\affiliation{NSF NOIRLab, 950 N. Cherry Ave., Tucson, AZ 85719, USA}

\author[0000-0002-5665-7912]{Biprateep~Dey}
\affiliation{Department of Astronomy \& Astrophysics, University of Toronto, Toronto, ON M5S 3H4, Canada}
\affiliation{Department of Physics \& Astronomy and Pittsburgh Particle Physics, Astrophysics, and Cosmology Center (PITT PACC), University of Pittsburgh, 3941 O'Hara Street, Pittsburgh, PA 15260, USA}

\author{P.~Doel}
\affiliation{Department of Physics \& Astronomy, University College London, Gower Street, London, WC1E 6BT, UK}

\author[0000-0002-3033-7312]{A.~Font-Ribera}
\affiliation{Institut de F\'{i}sica d’Altes Energies (IFAE), The Barcelona Institute of Science and Technology, Edifici Cn, Campus UAB, 08193, Bellaterra (Barcelona), Spain}

\author[0000-0002-2890-3725]{J.~E.~Forero-Romero}
\affiliation{Departamento de F\'isica, Universidad de los Andes, Cra. 1 No. 18A-10, Edificio Ip, CP 111711, Bogot\'a, Colombia}
\affiliation{Observatorio Astron\'omico, Universidad de los Andes, Cra. 1 No. 18A-10, Edificio H, CP 111711 Bogot\'a, Colombia}

\author{E.~Gaztañaga}
\affiliation{Institut d'Estudis Espacials de Catalunya (IEEC), c/ Esteve Terradas 1, Edifici RDIT, Campus PMT-UPC, 08860 Castelldefels, Spain}
\affiliation{Institute of Cosmology and Gravitation, University of Portsmouth, Dennis Sciama Building, Portsmouth, PO1 3FX, UK}
\affiliation{Institute of Space Sciences, ICE-CSIC, Campus UAB, Carrer de Can Magrans s/n, 08913 Bellaterra, Barcelona, Spain}

\author[0000-0003-3142-233X]{S.~Gontcho A Gontcho}
\affiliation{Lawrence Berkeley National Laboratory, 1 Cyclotron Road, Berkeley, CA 94720, USA}

\author{G.~Gutierrez}
\affiliation{Fermi National Accelerator Laboratory, PO Box 500, Batavia, IL 60510, USA}

\author[0000-0001-9822-6793]{J.~Guy}
\affiliation{Lawrence Berkeley National Laboratory, 1 Cyclotron Road, Berkeley, CA 94720, USA}

\author[0000-0003-1197-0902]{C.~Hahn}
\affiliation{Steward Observatory, University of Arizona, 933 N. Cherry Avenue, Tucson, AZ 85721, USA}
\affiliation{Steward Observatory, University of Arizona, 933 N. Cherry Avenue, Tucson, AZ 85721, USA}

\author[0000-0002-6550-2023]{K.~Honscheid}
\affiliation{Center for Cosmology and AstroParticle Physics, The Ohio State University, 191 West Woodruff Avenue, Columbus, OH 43210, USA}
\affiliation{Department of Physics, The Ohio State University, 191 West Woodruff Avenue, Columbus, OH 43210, USA}
\affiliation{The Ohio State University, Columbus, 43210 OH, USA}

\author[0000-0002-6024-466X]{M.~Ishak}
\affiliation{Department of Physics, The University of Texas at Dallas, 800 W. Campbell Rd., Richardson, TX 75080, USA}

\author[0000-0001-8528-3473]{J.~Jimenez}
\affiliation{Institut de F\'{i}sica d’Altes Energies (IFAE), The Barcelona Institute of Science and Technology, Edifici Cn, Campus UAB, 08193, Bellaterra (Barcelona), Spain}

\author{R.~Kehoe}
\affiliation{Department of Physics, Southern Methodist University, 3215 Daniel Avenue, Dallas, TX 75275, USA}

\author[0000-0002-8828-5463]{D.~Kirkby}
\affiliation{Department of Physics and Astronomy, University of California, Irvine, 92697, USA}

\author[0000-0003-3510-7134]{T.~Kisner}
\affiliation{Lawrence Berkeley National Laboratory, 1 Cyclotron Road, Berkeley, CA 94720, USA}

\author[0000-0001-6356-7424]{A.~Kremin}
\affiliation{Lawrence Berkeley National Laboratory, 1 Cyclotron Road, Berkeley, CA 94720, USA}

\author[0000-0002-6731-9329]{C.~Lamman}
\affiliation{Center for Astrophysics $|$ Harvard \& Smithsonian, 60 Garden Street, Cambridge, MA 02138, USA}

\author[0000-0003-1838-8528]{M.~Landriau}
\affiliation{Lawrence Berkeley National Laboratory, 1 Cyclotron Road, Berkeley, CA 94720, USA}

\author[0000-0001-7178-8868]{L.~Le~Guillou}
\affiliation{Sorbonne Universit\'{e}, CNRS/IN2P3, Laboratoire de Physique Nucl\'{e}aire et de Hautes Energies (LPNHE), FR-75005 Paris, France}

\author[0000-0003-1887-1018]{M.~E.~Levi}
\affiliation{Lawrence Berkeley National Laboratory, 1 Cyclotron Road, Berkeley, CA 94720, USA}

\author[0000-0003-4962-8934]{M.~Manera}
\affiliation{Departament de F\'{i}sica, Serra H\'{u}nter, Universitat Aut\`{o}noma de Barcelona, 08193 Bellaterra (Barcelona), Spain}
\affiliation{Institut de F\'{i}sica d’Altes Energies (IFAE), The Barcelona Institute of Science and Technology, Edifici Cn, Campus UAB, 08193, Bellaterra (Barcelona), Spain}

\author[0000-0002-4279-4182]{P.~Martini}
\affiliation{Center for Cosmology and AstroParticle Physics, The Ohio State University, 191 West Woodruff Avenue, Columbus, OH 43210, USA}
\affiliation{Department of Astronomy, The Ohio State University, 4055 McPherson Laboratory, 140 W 18th Avenue, Columbus, OH 43210, USA}
\affiliation{The Ohio State University, Columbus, 43210 OH, USA}

\author[0000-0002-1125-7384]{A.~Meisner}
\affiliation{NSF NOIRLab, 950 N. Cherry Ave., Tucson, AZ 85719, USA}

\author{R.~Miquel}
\affiliation{Instituci\'{o} Catalana de Recerca i Estudis Avan\c{c}ats, Passeig de Llu\'{\i}s Companys, 23, 08010 Barcelona, Spain}
\affiliation{Institut de F\'{i}sica d’Altes Energies (IFAE), The Barcelona Institute of Science and Technology, Edifici Cn, Campus UAB, 08193, Bellaterra (Barcelona), Spain}

\author[0000-0002-2733-4559]{J.~Moustakas}
\affiliation{Department of Physics and Astronomy, Siena College, 515 Loudon Road, Loudonville, NY 12211, USA}

\author[0000-0001-9070-3102]{S.~Nadathur}
\affiliation{Institute of Cosmology and Gravitation, University of Portsmouth, Dennis Sciama Building, Portsmouth, PO1 3FX, UK}

\author[0000-0003-3188-784X]{N.~Palanque-Delabrouille}
\affiliation{IRFU, CEA, Universit\'{e} Paris-Saclay, F-91191 Gif-sur-Yvette, France}
\affiliation{Lawrence Berkeley National Laboratory, 1 Cyclotron Road, Berkeley, CA 94720, USA}

\author[0000-0002-0644-5727]{W.~J.~Percival}
\affiliation{Department of Physics and Astronomy, University of Waterloo, 200 University Ave W, Waterloo, ON N2L 3G1, Canada}
\affiliation{Perimeter Institute for Theoretical Physics, 31 Caroline St. North, Waterloo, ON N2L 2Y5, Canada}
\affiliation{Waterloo Centre for Astrophysics, University of Waterloo, 200 University Ave W, Waterloo, ON N2L 3G1, Canada}

\author[0000-0001-7145-8674]{F.~Prada}
\affiliation{Instituto de Astrof\'{i}sica de Andaluc\'{i}a (CSIC), Glorieta de la Astronom\'{i}a, s/n, E-18008 Granada, Spain}

\author[0000-0001-6979-0125]{I.~P\'erez-R\`afols}
\affiliation{Departament de F\'isica, EEBE, Universitat Polit\`ecnica de Catalunya, c/Eduard Maristany 10, 08930 Barcelona, Spain}

\author{G.~Rossi}
\affiliation{Department of Physics and Astronomy, Sejong University, 209 Neungdong-ro, Gwangjin-gu, Seoul 05006, Republic of Korea}

\author[0000-0002-9646-8198]{E.~Sanchez}
\affiliation{CIEMAT, Avenida Complutense 40, E-28040 Madrid, Spain}

\author{M.~Schubnell}
\affiliation{Department of Physics, University of Michigan, 450 Church Street, Ann Arbor, MI 48109, USA}
\affiliation{University of Michigan, 500 S. State Street, Ann Arbor, MI 48109, USA}

\author[0000-0002-3461-0320]{J.~Silber}
\affiliation{Lawrence Berkeley National Laboratory, 1 Cyclotron Road, Berkeley, CA 94720, USA}

\author{D.~Sprayberry}
\affiliation{NSF NOIRLab, 950 N. Cherry Ave., Tucson, AZ 85719, USA}

\author[0000-0003-1704-0781]{G.~Tarl\'{e}}
\affiliation{University of Michigan, 500 S. State Street, Ann Arbor, MI 48109, USA}

\author{B.~A.~Weaver}
\affiliation{NSF NOIRLab, 950 N. Cherry Ave., Tucson, AZ 85719, USA}

\author[0000-0003-2229-011X]{R.~H.~Wechsler}
\affiliation{Kavli Institute for Particle Astrophysics and Cosmology, Stanford University, Menlo Park, CA 94305, USA}
\affiliation{Physics Department, Stanford University, Stanford, CA 93405, USA}
\affiliation{SLAC National Accelerator Laboratory, 2575 Sand Hill Road, Menlo Park, CA 94025, USA}

\author[0000-0001-5381-4372]{R.~Zhou}
\affiliation{Lawrence Berkeley National Laboratory, 1 Cyclotron Road, Berkeley, CA 94720, USA}

\author[0000-0002-6684-3997]{H.~Zou}
\affiliation{National Astronomical Observatories, Chinese Academy of Sciences, A20 Datun Road, Chaoyang District, Beijing, 100101, P.~R.~China}